%% file: main.tex
\documentclass[reprint,aps,superscriptaddress,
amsmath,amsfonts,amssymb,prb,floatfix]{revtex4-2}

\usepackage{graphicx}
\usepackage{bm,braket}
\usepackage{xcolor}
\usepackage{color}
\usepackage{mathrsfs}
\usepackage{mathtools}
\usepackage[colorlinks=true,linkcolor=blue,citecolor=blue,urlcolor=blue]{hyperref}
\usepackage{orcidlink}

\begin{document}

\title{Generalized Bloch-like ground states in distinct electron-site configurations derived from the same quasiperiodic structure}

\author{Yuki Yamamoto\,\orcidlink{0009-0009-5018-6126}}
\email{yamamoto.yuki.5p3@ecs.osaka-u.ac.jp}
\affiliation{
\mbox{Department of Physics, Graduate School of Science, The University of Osaka, Toyonaka, Osaka 560-0043, Japan}}
\author{Nayuta Takemori\,\orcidlink{0000-0001-9767-8632}}
\email{nayuta.takemori.sci@osaka-u.ac.jp}
\affiliation{
\mbox{Department of Physics, Graduate School of Science, The University of Osaka, Toyonaka, Osaka 560-0043, Japan}}
\affiliation{
\mbox{Center for Quantum Information and Quantum Biology, The University of Osaka, Toyonaka, Osaka 560-0043, Japan
}}
\affiliation{
\mbox{Forefront Research Center, The University of Osaka, Toyonaka, Osaka 560-0043, Japan 
}}

\begin{abstract}
Sutherland-Kalugin-Katz (SKK) states have been proposed as generalized Bloch-like states in quasiperiodic systems where the absence of translational symmetry prevents the application of Bloch's theorem. However, previous studies have largely relied on ansatz-based arguments, and a systematic procedure for constructing the SKK state from a specified Hamiltonian has remained unclear. In this work, we develop a constructive approach based on a generalized Fourier transform of creation operators in tight-binding models. We then apply this framework to vertex and dual models on the Ammann-Beenker (AB) tiling that stem from the same quasiperiodicity but differ in their electron-site configurations. Our construction reproduces the previously reported SKK state in the ground state of the vertex model and reveals a corresponding SKK state in the ground state of the dual model with high fidelity. The structural correspondence between the two SKK states indicates that such states are not tied to a particular choice of electron sites and reflect the underlying quasiperiodicity of the AB tiling.
\end{abstract}

\maketitle

\section{Introduction}
Quasicrystals possess long-range order without translational symmetry. Since their discovery in Al-Mn alloys~\cite{Shechtman1984}, they have been understood as ordered structures that can exhibit sharp Bragg peaks while possessing rotational symmetries forbidden in periodic crystals~\cite{Levine1984}. The absence of translational symmetry makes the electronic problem in quasicrystals fundamentally different from that in periodic crystals, where Bloch's theorem labels single-particle eigenstates by crystal momentum and reduces the Schr\"odinger equation to a problem within a unit cell~\cite{Bloch1928,Ashcroft1976}. In quasicrystals, however, crystal momentum is not a good quantum number, and Bloch's theorem cannot be applied.

Quasiperiodic systems can also host unconventional single-particle states, such as critical states that are neither completely localized nor extended~\cite{Kohmoto1983,Ostlund1983,Niu1986,Tsunetsugu1986,Kohmoto1987,Tokihiro1988,Tsunetsugu1991}. These examples show that quasiperiodicity can strongly affect single-particle wave functions. Therefore, constructing a useful description of single-particle eigenstates is an important step toward understanding electronic properties that are specific to quasiperiodic systems.

A promising candidate for such a description is the class of Sutherland-Kalugin-Katz (SKK) states introduced by Kalugin and Katz~\cite{Kalugin2014}, building on the work of Sutherland~\cite{Sutherland1986}. They were proposed as Bloch-like states that may describe electronic states in quasiperiodic systems. Similar to a Bloch state, the wave function is expressed as the product of a prefactor and an exponential factor. Certain electronic states in specific quasiperiodic tight-binding models have been shown to be well described by this form~\cite{Kalugin2014,Mace2017,Jeon2021}. However, previous studies mainly treated the SKK state as an ansatz motivated by structural features of quasiperiodicity and fixed its parameters by comparison with numerical eigenstates. Thus, although the usefulness of SKK states was demonstrated, a systematic procedure for constructing them from a specified Hamiltonian remained unclear.

In this work, we develop a constructive approach based on a generalized Fourier transform of creation operators. This approach is motivated by the role of the ordinary Fourier transform in Bloch's theorem, where a periodic Hamiltonian is decomposed into momentum sectors. For quasiperiodic systems, we look for a generalized Fourier transform that makes the action of the Hamiltonian closed within the subspace generated by the transformed operators. The Hamiltonian restricted to this closed subspace gives an effective eigenvalue problem, from which the corresponding SKK states are constructed. In this sense, within the present framework, the SKK states are formulated through this effective eigenvalue problem, rather than being merely treated as an ansatz.

We apply this construction to tight-binding models on the Ammann-Beenker (AB) tiling \cite{Beenker1982,Ammann1992}, a standard two-dimensional quasiperiodic structure with eightfold rotational symmetry. We then consider the vertex and dual models on this tiling that stem from the same quasiperiodicity but differ in their electron-site configurations. The vertex model is used as a benchmark because its ground state is known to exhibit an SKK state \cite{Kalugin2014,Mace2017}. The dual model serves as the main application of our construction in this work.

We show that our construction reproduces the known SKK state in the ground state of the vertex model and demonstrates that the ground state of the dual model is also described by an SKK state with high fidelity. We further find a structural feature common to the two SKK states. These results indicate that such states are not restricted to a particular electron-site configuration. Moreover, this common structural feature can be regarded as reflecting the underlying quasiperiodicity of the AB tiling.

The rest of this paper is organized as follows.
In Sec.~\ref{sec:model}, we introduce the models used in this work and describe our theoretical framework for constructing SKK states.
In Sec.~\ref{sec:result}, we present the numerical results for the constructed SKK states in the vertex and dual models, and discuss the comparison between their structures.
Finally, Sec.~\ref{sec:conclusion} summarizes our results.

\section{Model and Method}
\label{sec:model} 
In this section, we first review the AB tiling and its square approximants in Sec.~\ref{subsec:ABtiling}. We then define the tight-binding Hamiltonians and the two distinct electron-site configurations on the AB tiling in Sec.~\ref{subsec:Tight-binding}. The mathematical construction of the SKK state underlying our framework, including the generalized Fourier transform, is presented in Sec.~\ref{subsec:framework}. Its application to the AB tiling is described in Sec.~\ref{subsec:framework2}.

\subsection{Ammann-Beenker tiling and its square approximants}
\label{subsec:ABtiling}
\begin{figure}[t]
\centering
\includegraphics[width=0.34\textwidth]{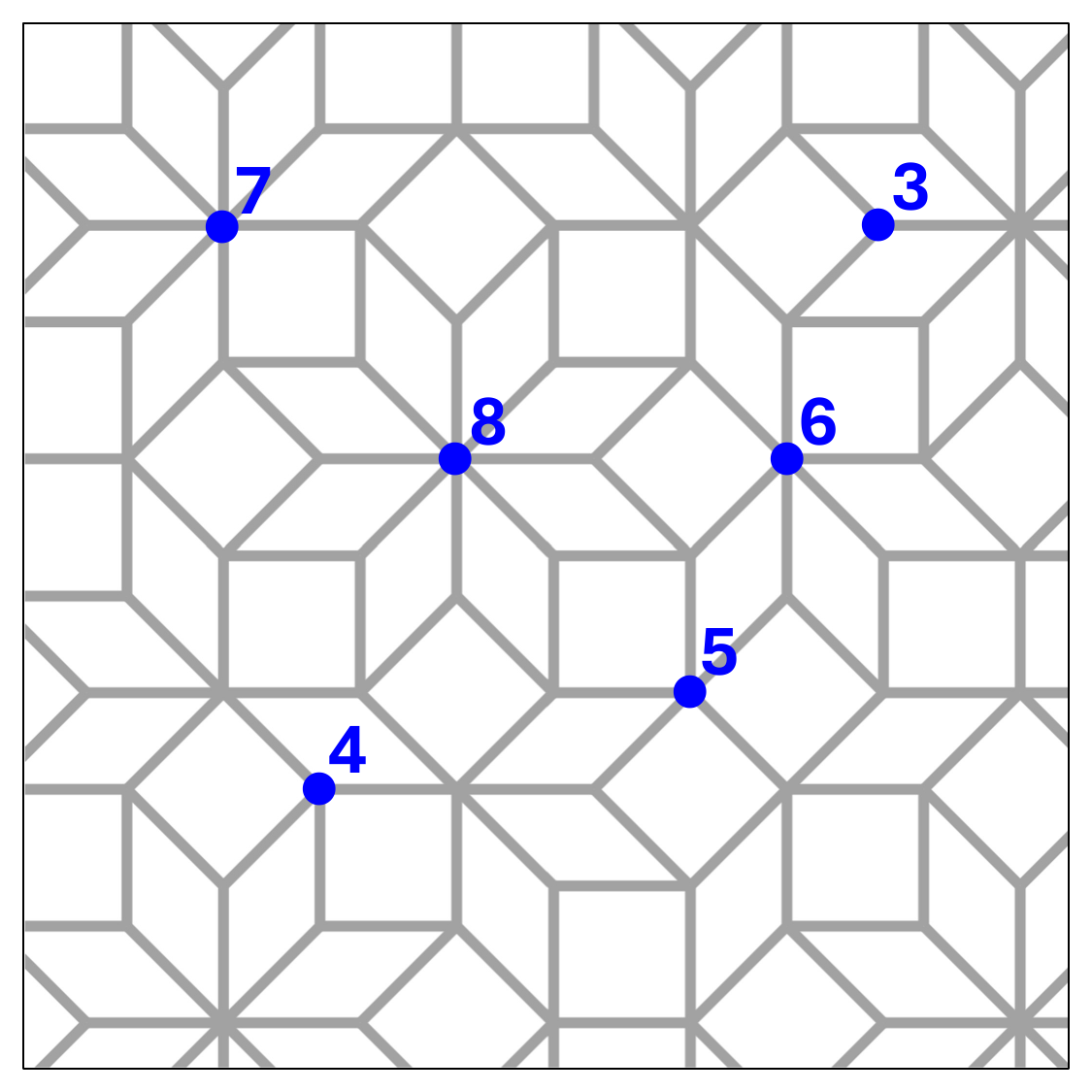}
\caption{Portion of the Ammann-Beenker tiling. Blue dots indicate representative vertices, and the adjacent numbers denote their coordination numbers, which vary from 3 to 8 depending on the position.}
\label{fig:ABtiling}
\end{figure}

We introduce the AB tiling, which uses squares and rhombi as its prototiles. In this quasiperiodic structure, the coordination number varies depending on the vertex location, taking values between 3 and 8, as shown in Fig.~\ref{fig:ABtiling}. The vertices of the AB tiling are generated via the cut-and-project method by projecting a four-dimensional hypercubic lattice onto a two-dimensional subspace \cite{Duneau1989, Jagannathan2024}.

Any lattice point in the four-dimensional space is described by a linear combination of the fundamental translation vectors $\{\bm{a}_\mu\}_{\mu=1}^4$ with integer coefficients.
This method operates by decomposing the four-dimensional space into two mutually orthogonal subspaces: the physical space $\mathcal{S}_{||}$ spanned by the orthonormal basis vectors $\bm{e}_1, \bm{e}_2$, and the perpendicular space $\mathcal{S}_{\perp}$ spanned by $\bm{e}_3, \bm{e}_4$.
Crucially, the AB tiling is not formed by simply projecting all four-dimensional lattice points onto $\mathcal S_{||}$. Instead, a specific selection criterion is applied. A lattice point is projected onto the physical space $\mathcal{S}_{||}$ only if its corresponding projection onto the perpendicular space $\mathcal{S}_{\perp}$ falls within a domain known as the acceptance window $W\subset \mathcal{S}_{\perp}$. For the AB tiling, $W$ is mathematically defined as the projection of the half-open hypercube $(-0.5,0.5]^4\subset\mathbb{R}^4$ onto $\mathcal{S}_{\perp}$, which results in a regular octagon.
The relationship connecting the fundamental vectors $\{\bm{a}_\mu\}_{\mu=1}^4$ to the orthonormal basis $\{\bm{e}_\mu\}_{\mu=1}^4$ is determined by the transformation matrix $C$, formulated as
\begin{align}
\label{eq:relationae} 
\begin{pmatrix}
\bm{a}_1 & \bm{a}_2 & \bm{a}_3 & \bm{a}_4
\end{pmatrix}
=
\begin{pmatrix}
\bm{e}_1 & \bm{e}_2 & \bm{e}_3 & \bm{e}_4
\end{pmatrix}
C,
\end{align}
where this matrix $C$ is explicitly defined by
\begin{align}
\label{eq:matrixC} 
C=\frac{1}{2}
\begin{pmatrix}
\sqrt{2} & 1 & 0 & -1 \\
0 & 1 & \sqrt{2} & 1 \\
\sqrt{2} & -1 & 0 & 1 \\
0 & 1 & -\sqrt{2} & 1
\end{pmatrix}.
\end{align}

\begin{figure*}[t]
\centering
\includegraphics[width=0.82\textwidth]{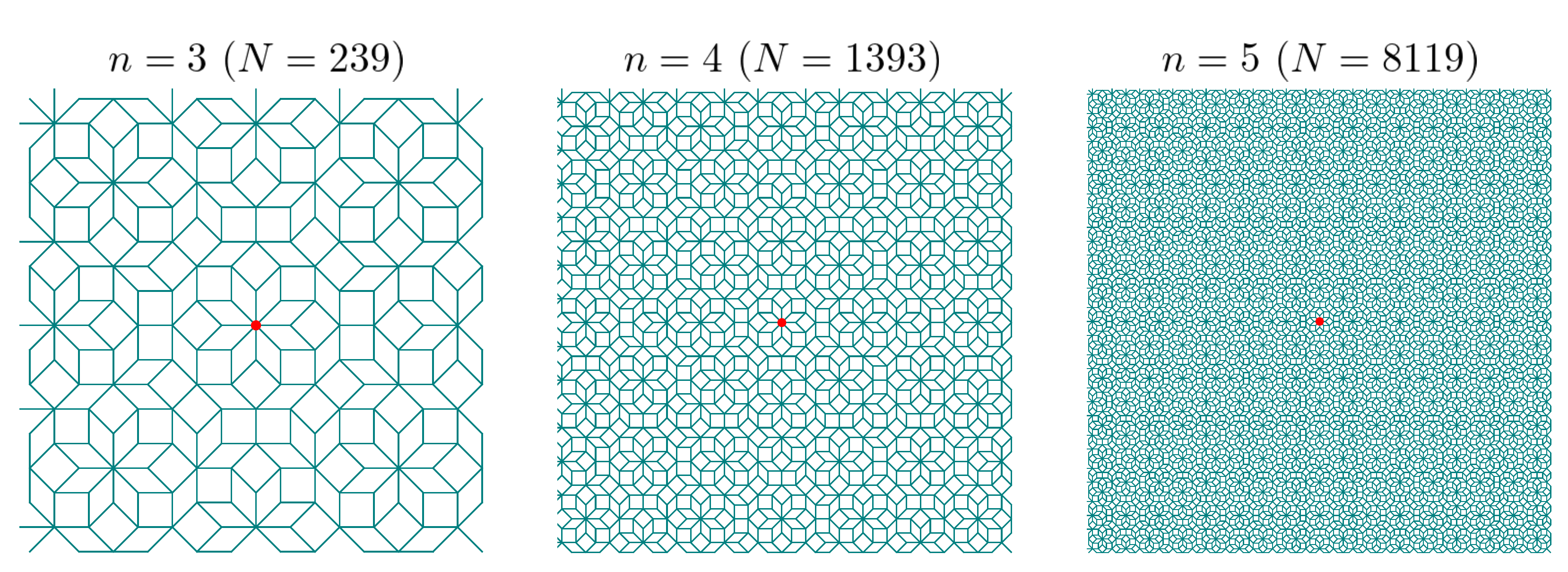}
\caption{Square approximants of generations $n=3,4,5$, with corresponding numbers of sites $N=239,1393,8119$. Red dot at the center marks the origin (eightfold rotational symmetry center). Unit cell size increases with $n$, providing a better approximation to the infinite quasiperiodic system.}
\label{fig:approximants}
\end{figure*}

In our numerical calculations, we adopt a finite square approximant to represent the infinite quasiperiodic tiling \cite{Jagannathan2024}. 
The transformation matrix corresponding to the $n$-th square approximant, denoted by $C_n$, is given by:
\begin{align}
\label{eq:matrixCn}
C_n=\frac{P_n}{2N_n} \begin{pmatrix} 
\sqrt{2}\alpha_n & \sqrt{2} & 0 & -\sqrt{2} \\ 
0 & \sqrt{2} & \sqrt{2}\alpha_n & \sqrt{2} \\
2 & -\alpha_n & 0 & \alpha_n \\
0 & \alpha_n & -2 & \alpha_n
\end{pmatrix}.
\end{align}
Here, $\alpha_n = 1 + P_{n-1}/P_n$, where $P_n$ is the $n$-th Pell number, and $N_n$ is given by $2N_n^2 = 4P_n^2 + (-1)^n$.
In the limit of $n \to \infty$, we obtain the original transformation matrix $C$ as 
\begin{align}
\label{eq:limitCn}
\lim_{n\to\infty}C_n=C.
\end{align}
Based on this construction, the side length $L$ of the $n$-th square approximant is given by $L = \sqrt{2} N_n$. We display the unit cells of the square approximants for $n=3, 4, 5$ in Fig.~\ref{fig:approximants}. In the numerical calculations below, we use square approximants with $n=4,5,6,7$, whose numbers of sites in the unit cell are $N=1\,393,8\,119,47\,321,275\,807$, respectively.

\begin{figure}[t]
\centering
\includegraphics[width=0.466\textwidth]{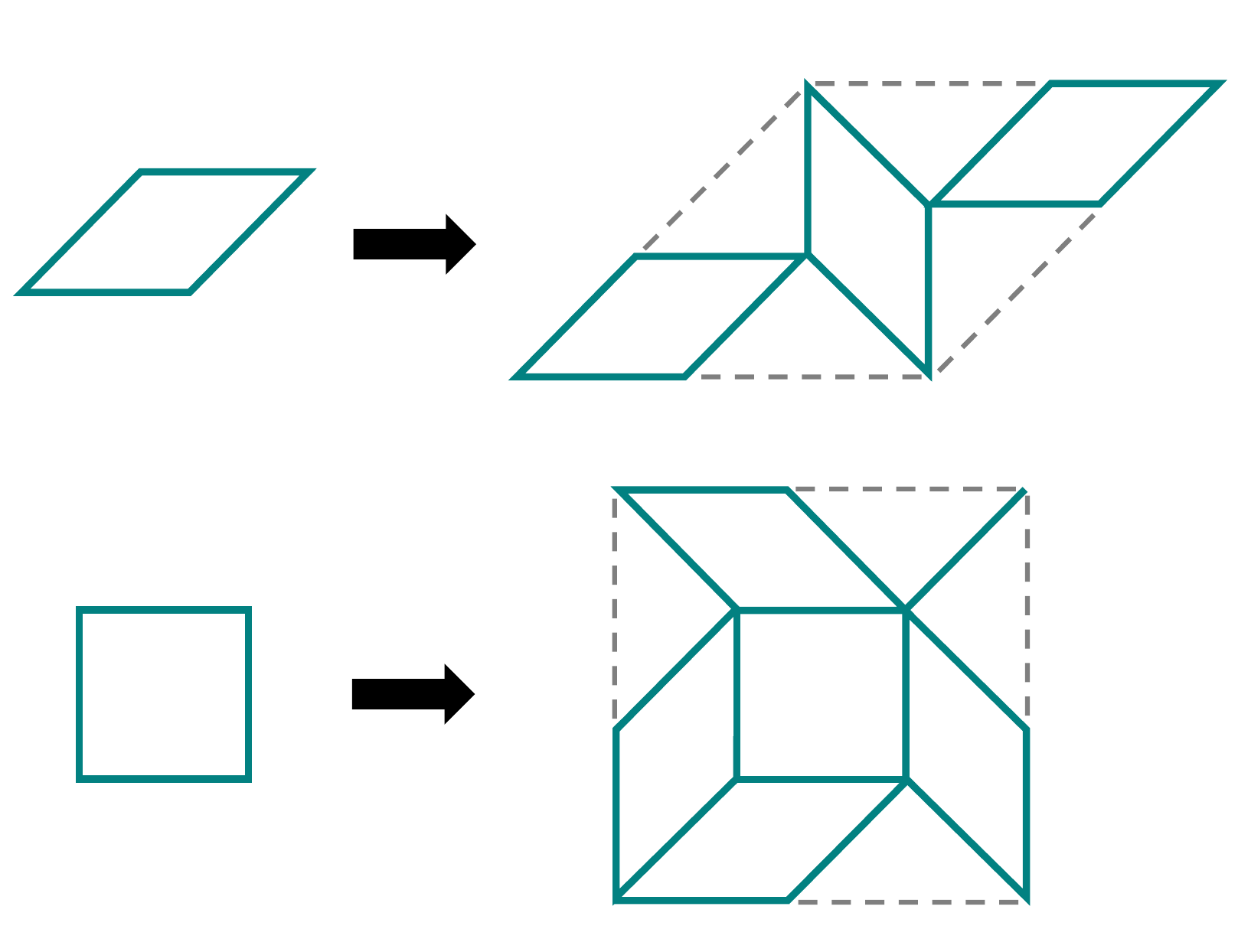}
\caption{Inflation rules for the Ammann-Beenker tiling. 
Each prototile is transformed into a patch composed of square and rhombic tiles.}
\label{fig:Inf_rule}
\end{figure}

We note that the AB tiling can also be constructed by successive transformation of the square and rhombic prototiles using inflation rules, as illustrated in Fig.~\ref{fig:Inf_rule}~\cite{Jagannathan2024}. Although the square approximants used in our numerical calculations are generated by the cut-and-project construction described above, the inflation rule will play an important role in the definition of the height function introduced later.

\subsection{Tight-binding models on quasicrystals}
\label{subsec:Tight-binding}
To investigate the electronic states on the AB tiling introduced above, we consider a tight-binding model on the AB tiling. The Hamiltonian is given by
\begin{align}
\label{eq:TBhamiltonian}
\hat{\mathscr{H}}=-\sum_{\langle i, j \rangle}t_{ij}(\hat{c}_i^\dagger \hat{c}_j+\hat{c}_j^\dagger \hat{c}_i)+\sum_i\left(\sum_{j\in N(i)}V_{ij}
\right)\hat{c}_i^\dagger \hat{c}_i.
\end{align}
Here, $\hat{c}_i^{(\dagger)}$ annihilates (creates) an electron at site $i$.
The notation $\langle i,j \rangle$ indicates nearest-neighbor pairs, and $N(i)$ denotes the set of nearest neighbors of site $i$.
Assuming a spherically symmetric potential, the hopping amplitude $t_{ij}$ and the onsite-potential contribution $V_{ij}$ depend only on the inter-site distance $r_{ij}=|\mathbf{r}_i - \mathbf{r}_j|$. Therefore, to explicitly determine these parameters, we must first define the spatial arrangement of the electron sites.

\begin{figure}[t]
\centering
\includegraphics[width=0.45\textwidth]{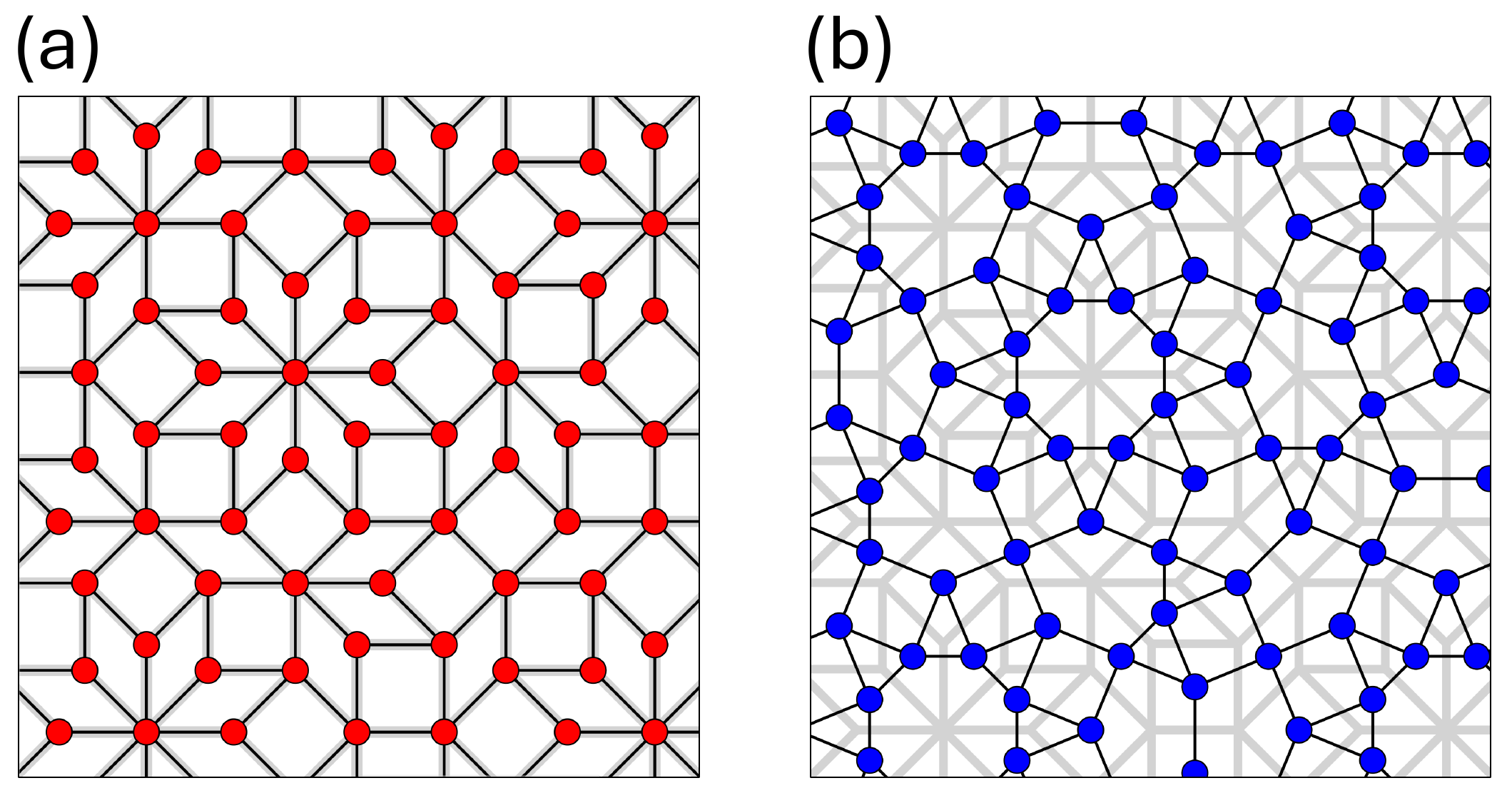}
\caption{Two lattice configurations derived from the Ammann-Beenker (AB) tiling. The underlying AB tiling is shown with light gray lines.
(a) Vertex model with electron sites (red dots) at tiling vertices and bonds (black lines) along tile edges. The bond length is constant and the coordination number $z_i$ varies from 3 to 8. 
(b) Dual model with electron sites (blue dots) at tile centers and bonds (black lines) connecting adjacent centers. Unlike the vertex model, the bond length is non-uniform, whereas the coordination number is fixed at $z_i=4$.}
\label{fig:VandD}
\end{figure}

In this study, we consider two distinct electron-site configurations: the vertex model (Fig.~\ref{fig:VandD}(a)) and the dual model (Fig.~\ref{fig:VandD}(b)). In the vertex model, the electron sites are positioned at the vertices of the tiling, with the bonds established along the edges of the constituent squares and rhombi. All bond lengths are identical, whereas the coordination number $z_i$ varies from 3 to 8. Consequently, the bond-dependent parameters become spatially uniform, with \(t_{ij}=t\) and \(V_{ij}=V\). The tight-binding Hamiltonian then takes the form
\begin{align}
\label{eq:TBhamiltonian2}
\hat{\mathscr{H}}=-\sum_{\langle i, j \rangle}t(\hat{c}_i^\dagger \hat{c}_j+\hat{c}_j^\dagger \hat{c}_i)+\sum_iVz_i\hat{c}_i^\dagger \hat{c}_i.
\end{align}
In our calculations for the vertex model, we fix the hopping amplitude by setting $t=1$ as the unit of energy, while treating the onsite potential $V$ as a continuous tuning parameter. 

In contrast, the dual model provides an alternative configuration where the electron sites are positioned at the centers of the constituent tiles. In this setup, bonds connect adjacent tiles across their shared edges. The distances between adjacent sites are non-uniform, depending on the specific pairing of adjacent tiles (e.g., square-square, square-rhombus, or rhombus-rhombus). Unlike the vertex model, however, the coordination number in the dual model constructed from the AB tiling is uniformly fixed at \(z_i=4\). This uniformity is a direct consequence of the underlying tiling geometry, reflecting the fact that all tiles have four edges. To capture this spatial dependence, we adopt exponentially decaying forms for the parameters, setting $t_{ij} = e^{-r_{ij}}$ and $V_{ij} = \lambda e^{-r_{ij}}$. This formulation ensures a constant ratio $V_{ij}/t_{ij} = \lambda$ for every bond. Accordingly, $\lambda$ serves as the continuous tuning parameter to modulate the potential landscape of the dual model. The present choice should be viewed as a representative Hamiltonian for the dual model controlled by the single parameter \(\lambda\), rather than as the most general tight-binding Hamiltonian with independently chosen hopping amplitudes and onsite potentials.

\subsection{Framework for eigenstates in quasiperiodic systems}
\label{subsec:framework}
\subsubsection{Sutherland-Kalugin-Katz state}
In periodic systems, the exact eigenstates of the tight-binding Hamiltonian are characterized by Bloch's theorem,
\begin{align}
\label{eq:Bloch_form}
\psi_{n,\mathbf{k}}^{\text{Bloch}}(\mathbf{r}) = u_{n,\mathbf{k}}(\mathbf{r}) e^{i\mathbf{k} \cdot \mathbf{r}},
\end{align}
where $n$ and $\mathbf{k}$ denote the band index and crystal momentum, respectively, and $u_{n,\mathbf{k}}(\mathbf{r})$ is a periodic function. However, the crystal momentum is ill-defined in quasicrystals, making the conventional approach inapplicable. 
 
To overcome this difficulty, Kalugin and Katz proposed the SKK state~\cite{Kalugin2014}, a generalized Bloch-like form for eigenstates in quasicrystals. This framework suggests that even in systems without periodicity, the exact eigenstates can occasionally be expressed in a form analogous to the Bloch function
\begin{align}
\label{eq:SKK_form}
\psi_{n,\Lambda}^{\text{SKK}}(\mathbf{r}) = \Psi_{n,\Lambda}(\mathbf{r}) e^{f_{\Lambda}(\mathbf{r})}.
\end{align}
Here, $\Psi_{n,\Lambda}(\mathbf{r})$ plays a role analogous to the periodic function $u_{n,\mathbf{k}}(\mathbf{r})$ in Eq.~(\ref{eq:Bloch_form}) and is therefore a quasiperiodic function, while $f_{\Lambda}(\mathbf{r})$ generalizes the phase factor associated with the generalized crystal momentum $\Lambda$. 

\subsubsection{Generalized Fourier transform}
In periodic tight-binding models, the framework of the SKK state reduces to Bloch's theorem. Crucially, Bloch's theorem is implemented in second quantization by Fourier transforming the creation and annihilation operators. This standard operation block-diagonalizes the Hamiltonian into independent sectors labeled by the crystal momentum. Motivated by this correspondence, we introduce a generalized Fourier transform to construct the SKK states in quasiperiodic tight-binding models. Using the exponential factor $e^{f_{\Lambda}(\mathbf{r})}$, we define the transformed operator as
\begin{align}
\label{eq:GeneralizedFourier2}
\hat{c}_{a,\Lambda}^{\dagger} &= \sum_{i \in S_a} e^{f_{\Lambda}(\mathbf{r}_i)} \hat{c}_i^{\dagger},
\end{align}
where the normalization constants are omitted for simplicity. Here, \(\{S_a\}\) represents a provisional classification of the sites into mutually disjoint subsets, which may be defined, for example, by local environments such as the coordination number.
Because the corresponding states have non-overlapping support, the states $\{\hat{c}_{a,\Lambda}^{\dagger}\ket{0} \}_a$ are mutually orthogonal and therefore linearly independent. The specific criteria for this classification, along with the appropriate form of the function $f_{\Lambda}(\mathbf{r})$, are determined by the closure condition discussed below.

\subsubsection{Closure condition and effective Hamiltonian}
As in Bloch's theorem, the transformed operators in Eq.~(\ref{eq:GeneralizedFourier2}) are useful when the Hamiltonian maps the subspace spanned by \(\{\hat c_{a,\Lambda}^{\dagger}\ket{0}\}_a\) into itself. We express this requirement as the closure condition
\begin{align}
\label{eq:commutator_condition}
[\hat{\mathscr{H}}, \hat{c}_{a,\Lambda}^{\dagger}] = \sum_{b} M_{ba} \hat{c}_{b,\Lambda}^{\dagger} \quad (M_{ba} \in \mathbb{C}).
\end{align}
When this condition is satisfied, the matrix \(M\) whose elements are \(M_{ab}\) represents the action of the Hamiltonian $\hat{\mathscr{H}}$ on the subspace labeled by \(\Lambda\).
The Schrödinger equation can be separated into an equation in this subspace. 
This leads to the reduced equation
\begin{align}
\label{eq:FiniteDim_EP}
\sum_{b} M_{ab} \Psi_{\Lambda}(b) = E \Psi_{\Lambda}(a).
\end{align}
Let \(\Psi_{m,\Lambda}(a)\) ($m = 0, 1, 2, \dots$) be the $m$-th solution of this eigenvalue problem.
For a site \(i\in S_a\), the corresponding real-space representation is
\begin{align}
\langle \mathbf{r}_i|\Psi_{m, \Lambda}\rangle = \Psi_{m,\Lambda}(a) e^{f_{\Lambda}(\mathbf{r}_i)}.
\end{align}
This has the form of an SKK state introduced in Eq.~(\ref{eq:SKK_form}). The prefactor \(\Psi_{m,\Lambda}(a)\) is defined on the site class \(S_a\), and therefore takes the same value for all sites belonging to the same class. If the closure condition is satisfied, the corresponding state is an eigenstate of the original Hamiltonian $\hat{\mathscr H}$ with eigenvalue $E$. The matrix $M$ therefore serves as an effective Hamiltonian (see Appendix~\ref{app:GFT_SKK}). The problem is reduced to finding a function \(f_{\Lambda}(\mathbf r_i)\) and a site classification \(\{S_a\}\) that simultaneously satisfy the closure condition.

In previous studies of the vertex model~\cite{Kalugin2014,Mace2017}, the SKK state was used as an ansatz motivated by structural features of quasiperiodicity, consisting of a prefactor associated with the local environment and an exponential factor $e^{f_{\Lambda}(\mathbf{r}_i)}$. The prefactor was then determined by comparison with numerically obtained ground states. In contrast,
in the present formulation, the prefactor \(\Psi_{m,\Lambda}(a)\) is obtained as an eigenvector of the effective Hamiltonian \(M\), once the exponent \(f_{\Lambda}(\mathbf r_i)\) and the site classification \(\{S_a\}\) are specified.

In the following, we first apply this formulation to the vertex model by using the structurally motivated exponent employed in previous studies and by choosing a site classification suitable for the closure condition. We then apply the same idea to the dual model, where the exponent and site classification are chosen in analogy with the vertex model. In practice, these classifications are truncated at a finite level to keep the problem numerically tractable. Specifically, we use seven classes for the vertex model and five classes for the dual model.

\subsection{Application to the Ammann-Beenker tiling}
\label{subsec:framework2}

To apply the above framework to the AB tiling, we evaluate the closure condition in Eq.~(\ref{eq:commutator_condition}) using the tight-binding model. Using the provisional classification \(\{S_a\}\), we define the following quantities for an arbitrary site \(i\in S_a\) and for each pair of classes \((a,b)\):
\begin{widetext}
\begin{align}
\label{eq:SKK_condition_vertex_local}
\mathcal{M}_{ab}^{\text{(v)}}(i)
&=
-\sum_{j \in S_b \cap N(i)}\ 
t e^{f_\Lambda(\mathbf{r}_j) - f_\Lambda(\mathbf{r}_i)}
+ Vz_i \delta_{ab}, \\
\label{eq:SKK_condition_dual_local}
\mathcal{M}_{ab}^{\text{(d)}}(i)
&=
-\sum_{j \in S_b \cap N(i)}\,
t_{ij} e^{f_\Lambda(\mathbf{r}_j) - f_\Lambda(\mathbf{r}_i)}
+
\left(
\sum_{k\in N(i)} V_{ik}
\right)
\delta_{ab}.
\end{align}
\end{widetext}
For the vertex and dual models, the closure condition is equivalent to
\begin{align}
\mathcal{M}_{ab}^{\text{(v)}}(i)&=M_{ab}^{\text{(v)}}
\qquad
\text{for all } i\in S_a,
\label{eq:SKK_condition_vertex}\\
\mathcal{M}_{ab}^{\text{(d)}}(i)&=M_{ab}^{\text{(d)}}
\qquad
\text{for all } i\in S_a.
\label{eq:SKK_condition_dual}
\end{align}
The matrices $M^{\text{(v)}}$ and $M^{\text{(d)}}$ then define the effective Hamiltonians for the vertex and dual models, respectively. The task is therefore to find an exponent \(f_\Lambda(\mathbf r)\) and a site classification \(\{S_a\}\) such that the conditions in Eqs.~\eqref{eq:SKK_condition_vertex} and \eqref{eq:SKK_condition_dual} are satisfied. The detailed derivation of these conditions is provided in Appendix~\ref{app:Cc_AB}.

The vertex model is treated first as a benchmark because its ground state is known from previous studies~\cite{Kalugin2014,Mace2017} to be well
described by an SKK state. The dual model is then considered as the main new application of this work. Its construction is developed in analogy with the SKK state of the vertex model ground state, and yields a new candidate for an SKK state in a model for which such a state has not been reported previously.

\subsubsection{Application to the vertex model}
\label{subsubsec:vertex_skk}
\begin{figure}[t]
\centering
\includegraphics[width=0.46\textwidth]{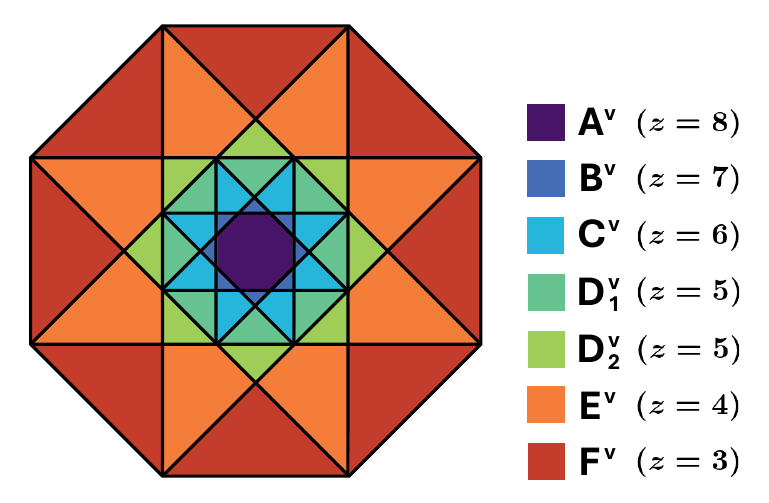}
\caption{Partition of the acceptance window \(W\subset \mathcal{S}_{\perp}\) into the seven site classes \(\mathrm{A^{v}},\mathrm{B^{v}},\mathrm{C^{v}},\mathrm{D^{v}_1},\mathrm{D^{v}_2},\mathrm{E^{v}}\) and \(\mathrm{F^{v}}\). In the cut-and-project construction, each vertex of the Ammann-Beenker tiling is associated with a point in \(W\), and the colored region containing this point determines its class. The coordination number \(z\) is constant within each class and is indicated in the legend.}
\label{fig:perpendicular}
\end{figure}

Here, we first recall the expression for the SKK state used as an ansatz in previous studies of the vertex model~\cite{Kalugin2014,Mace2017}, and then show how the same expression is obtained within our framework. In the cut-and-project description introduced in Sec.~\ref{subsec:ABtiling}, each vertex is associated with a point in the perpendicular space \(\mathcal{S}_{\perp}\). In previous studies, vertices were classified according to the region of the acceptance window \(W\subset \mathcal{S}_{\perp}\) that contains the corresponding point in the perpendicular space. This gives seven site classes labeled by
\begin{align}
a \in \{\mathrm{A^{v}},\mathrm{B^{v}},\mathrm{C^{v}},\mathrm{D^{v}_1},\mathrm{D^{v}_2},\mathrm{E^{v}},\mathrm{F^{v}}\},
\end{align}
as shown in Fig.~\ref{fig:perpendicular}. Most of these classes are distinguished by the coordination number of the vertex. The exception is the five-coordinated vertices, which are divided into two classes, \(\mathrm{D^{v}_1}\) and \(\mathrm{D^{v}_2}\).

\begin{figure}[!t]
\centering
\includegraphics[width=0.4\textwidth]{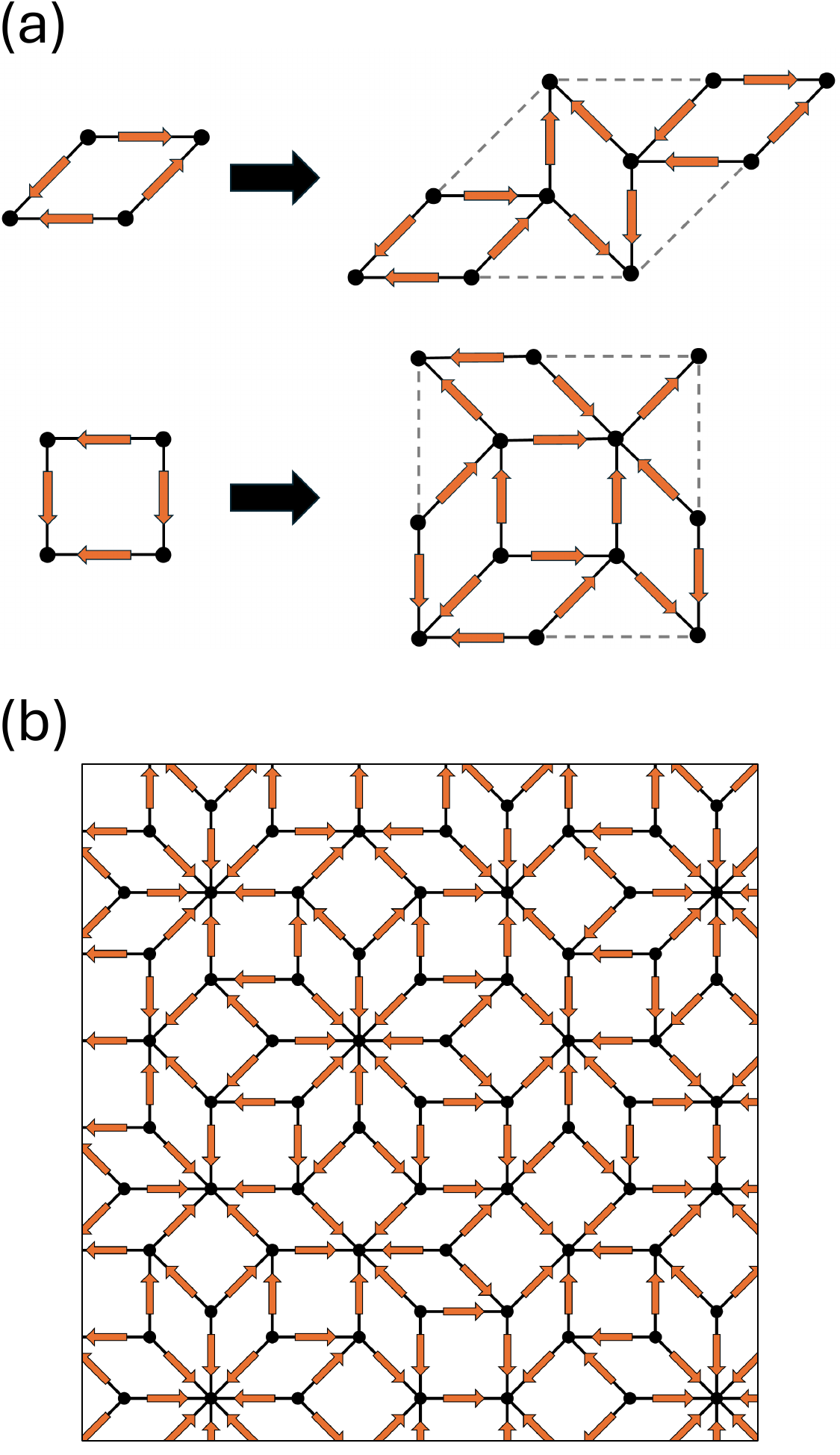}
\caption{(a) Tile inflation rules and arrow decorations corresponding to matching rules for the Ammann-Beenker tiling. (b) A globally curl-free arrow field generated on a larger patch of the tiling by applying the inflation rules shown in (a).}
\label{fig:arrows}
\end{figure}

For a site \(i\in S_a\), the ground state of the vertex model was numerically shown to be well described by
\begin{align}
\label{eq:vertex_SKK_state}
\psi(\mathbf{r}_i) = C(a)e^{\kappa h(\mathbf{r}_i)} ,
\end{align}
where \(C(a)\) is a prefactor depending only on the site class, \(\kappa\) is a real constant, and \(h(\mathbf{r}_i)\) is a height function~\cite{Kalugin2014,Mace2017}. The height function is obtained by integrating an arrow field that represents the matching rule of the AB tiling. This arrow field can be generated together with the tiling by applying the inflation rules, as shown in Fig.~\ref{fig:arrows}(a).
The value of the height function at a vertex \(i\) is defined as
\begin{equation}
\label{eq:height_integral}
\begin{aligned}
h(\mathbf{r}_i) &= \sum_{\langle j, k \rangle \in \Gamma_i} \sigma_{jk}, \\
\sigma_{jk} &= \begin{cases} 
+1 & (\text{if } j \to k \text{ is parallel to the arrow}) \\ 
-1 & (\text{if } j \to k \text{ is anti-parallel}) 
\end{cases}.
\end{aligned}
\end{equation}
Here, \(\Gamma_i\) is an arbitrary path on the edges of the tiling from a reference site to \(i\), and the sum runs over the nearest-neighbor steps \(\langle j,k\rangle\) along this path. Because the arrow field is curl-free, as shown in Fig.~\ref{fig:arrows}(b), the value of \(h(\mathbf{r}_i)\) is independent of the chosen path. Furthermore, changing the reference vertex shifts the entire height function by a constant, which is absorbed into the normalization constant.

To show how this expression is obtained within our framework, we first choose the exponent of the generalized Fourier factor as
\begin{align}
\label{eq:vertex_exponent}
f_\Lambda(\mathbf{r}_i)=\kappa h(\mathbf{r}_i).
\end{align}
For nearest-neighbor vertices \(i\) and \(j\), the height difference $h(\bm r_j)-h(\bm r_i)$ is determined by the arrow on the edge connecting them. Therefore, the factor
\begin{align}
e^{f_\Lambda(\mathbf{r}_j)-f_\Lambda(\mathbf{r}_i)}=e^{\kappa[h(\mathbf{r}_j)-h(\mathbf{r}_i)]}
\end{align}
is controlled by the local arrow configuration. To examine the closure condition, we substitute this exponent into Eq.~\eqref{eq:SKK_condition_vertex_local} and analyze the site dependence of \(\mathcal M_{ab}^{\text{(v)}}(i)\). We find that 
\(\mathcal M_{ab}^{\text{(v)}}(i)\) is determined by the coordination number, the classes of neighboring sites, and the local arrow configuration around the site \(i\). 

\begin{figure*}[t]
\centering
\includegraphics[width=0.78\textwidth]{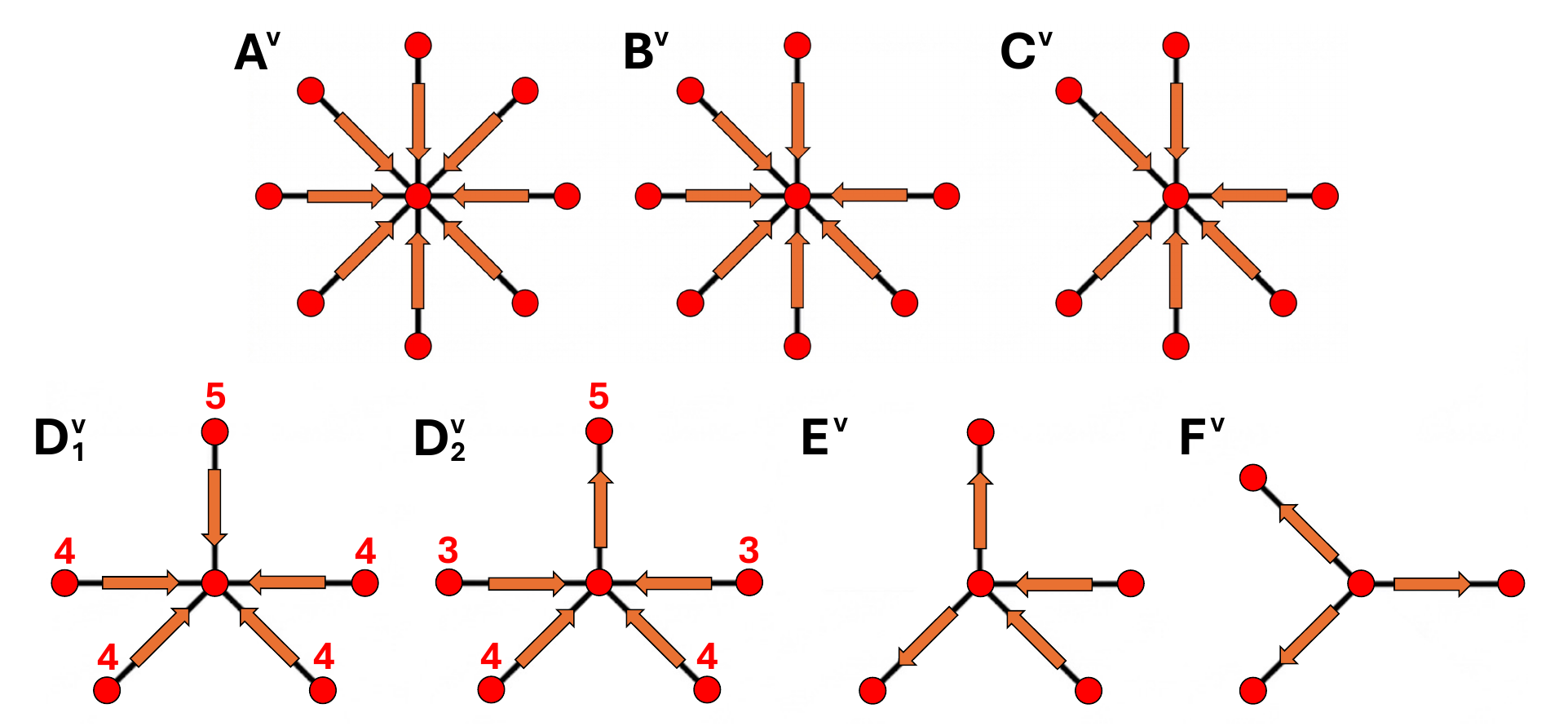}
\caption{Classification of the seven classes (\(\mathrm{A^{v}},\mathrm{B^{v}},\mathrm{C^{v}},\mathrm{D^{v}_1},\mathrm{D^{v}_2},\mathrm{E^{v}}\) and \(\mathrm{F^{v}}\)) in the vertex model. While most types are uniquely identified by their coordination number, \(\mathrm{D^{v}_1}\) and \(\mathrm{D^{v}_2}\) both have a coordination number five. They are distinguished by the coordination numbers of their neighboring sites: \(\mathrm{D^{v}_2}\) has a nearest-neighbor site with coordination number three, whereas \(\mathrm{D^{v}_1}\) does not. This difference reverses the direction of a single arrow between the two types.}
\label{fig:vertex_types}
\end{figure*}

The seven classes introduced above distinguish sites according to their coordination number and local arrow configuration, as shown in Fig.~\ref{fig:vertex_types}. Although this classification captures the local information relevant to the closure condition, it does not determine $\mathcal M_{ab}^{\text{(v)}}(i)$ uniquely, since the pairing between neighboring-site classes and the corresponding height differences is not always fixed. An exact closure condition would therefore require a further refinement of the site classification. In the present work, however, we retain this finite seven-class description, which yields a finite-dimensional effective Hamiltonian (Appendix~\ref{app:PraCons_M}).


As a result, the problem is reduced to the eigenvalue problem of the seven-dimensional effective Hamiltonian $M^{\text{(v)}}$ for each fixed $\kappa$. For a site \(i\in S_a\), the $m$-th wave function is given by
\begin{align}
\label{eq:vertex_wavefunction}
\psi_{m, \Lambda}(\mathbf r_i)=\Psi_{m, \Lambda}(a)e^{\kappa h(\mathbf r_i)}.
\end{align}
For the lowest eigenvector \(\psi_{0,\Lambda}(\mathbf r)\), this expression has the same structure as the ground state in Eq.~\eqref{eq:vertex_SKK_state}, with the class-dependent prefactor \(C(a)\) replaced by the eigenvector component \(\Psi_{0,\Lambda}(a)\). In this sense, the known SKK state of the vertex model is obtained within our framework from the effective eigenvalue problem in the present seven-class description. The other eigenvectors also generate wave functions of the form in Eq.~\eqref{eq:vertex_wavefunction}, although this finite-class approximation alone does not guarantee that they are eigenstates of the full tight-binding Hamiltonian.

The construction above is formulated in terms of the height function on the quasiperiodic AB tiling. To compare the resulting states with numerical eigenstates, however, it must be adapted to finite square approximants with periodic boundary conditions. Although these approximants are not generated directly by inflation (Sec.~\ref{subsec:ABtiling}), a height function can still be defined by decorating each square and rhombus with the arrow pattern of the corresponding prototile shown in Fig.~\ref{fig:arrows}(a), which yields a unique curl-free arrow field. The resulting height function is generally not periodic on the approximant unit cell and would therefore produce an artificial discontinuity across the boundary under periodic boundary conditions. To avoid this issue, we replace $h(\mathbf r_i)$ by a periodic height function $\tilde h(\mathbf r_i)$, constructed as described in Appendix~\ref{app:periodic_height}.

This correction has a consequence for the finite-class approximation. For the corrected height function, the difference \(\tilde h(\mathbf r_j)-\tilde h(\mathbf r_i)\) across a nearest-neighbor edge is not determined only by the arrow on that edge. As a result, the local data entering \(\mathcal M_{ab}^{\text{(v)}}(i)\) are modified in the square approximant. The construction of the finite-dimensional effective Hamiltonian on the square approximants, including this effect, is described in Appendix~\ref{app:PraCons_apprM}.

The imposed periodic boundary condition also restricts the values of \(\kappa\) that can give eigenstates of the approximants. Therefore, in Sec.~\ref{sec:result}, we treat \(\kappa\) as a variational parameter in the numerical construction of the SKK state within our framework. The optimal value of \(\kappa\) is determined by comparing the SKK state constructed from the lowest eigenvector of \(M^{\text{(v)}}\) with the numerical ground state of the full tight-binding Hamiltonian.

\subsubsection{Application to the dual model}
\label{subsubsec:dual_skk}
\begin{figure*}[t]
\centering
\includegraphics[width=0.78\textwidth]{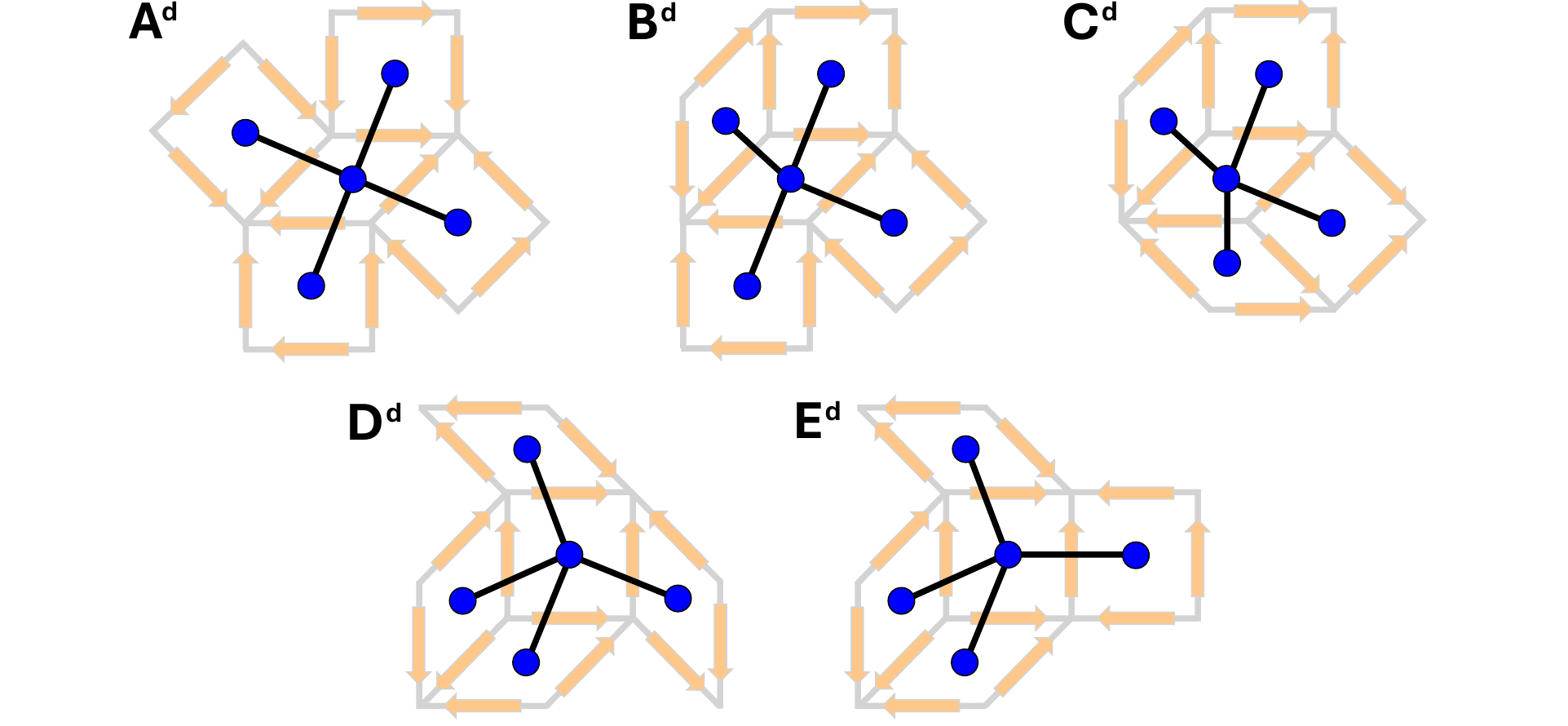}
\caption{Classification of the five classes (\(\mathrm{A^{d}},\mathrm{B^{d}},\mathrm{C^{d}},\mathrm{D^{d}}\) and \(\mathrm{E^{d}}\)) in the dual model. Blue circles denote dual sites located at tile centers, connected by black bonds. The constituent tiles and arrow field are indicated by gray lines and orange arrows, respectively. The types are therefore distinguished by the shape of their central tile (a rhombus for \(\mathrm{A^{d}}\), \(\mathrm{B^{d}}\), and \(\mathrm{C^{d}}\); a square for \(\mathrm{D^{d}}\) and \(\mathrm{E^{d}}\)) and the arrangement of their four neighboring tiles.}
\label{fig:dual_vertex_types}
\end{figure*}

We next consider the closure condition for the dual model in Eq.~\eqref{eq:SKK_condition_dual}. A site \(i\) in the dual model is located at the center of either a square or a rhombic tile of the AB tiling. We define a dual height function $h_{\rm d}(\mathbf r_i)$ by averaging the height function $h(\mathbf{r}_i)$ defined above over the four vertices of the corresponding tile,
\begin{align}
\label{eq:height_dual}
h_{\rm d}(\mathbf r_i)=\frac{1}{4}\sum_{v\in\mathcal T(i)}h(\mathbf r_v),
\end{align}
where \(\mathcal T(i)\) denotes the set of the four vertices of the tile whose center is the dual site \(i\). In analogy with the vertex model, we choose the exponent of the generalized Fourier factor for the dual model as
\begin{align}
\label{eq:dual_exponent}
f_\Lambda(\mathbf r_i)
=
\kappa' h_{\rm d}(\mathbf r_i),
\end{align}
where \(\kappa'\) is a real constant. To examine the closure condition, we substitute this exponent into Eq.~\eqref{eq:SKK_condition_dual_local} and analyze the site dependence of \(\mathcal M_{ab}^{\text{(d)}}(i)\). We find that \(\mathcal M_{ab}^{\text{(d)}}(i)\) is determined by the classes of neighboring dual sites, the bond lengths between adjacent dual sites, and the dual height differences \(h_{\rm d}(\mathbf r_j)-h_{\rm d}(\mathbf r_i)\). The bond lengths are fixed by the shapes of the adjacent tiles, while the dual height differences are determined by the arrow configuration on the edges of the tile containing the dual site \(i\) and the neighboring tiles sharing an edge with it. Thus, both the bond lengths and the relevant arrow configurations are determined by the local tile configuration around the dual site. We therefore classify the dual sites according to the local tile configurations. This leads to five classes,
\begin{align}
a\in\{\mathrm{A^{d}},\mathrm{B^{d}},\mathrm{C^{d}},\mathrm{D^{d}},\mathrm{E^{d}}\},
\end{align}
as shown in Fig.~\ref{fig:dual_vertex_types}. The classes \(\mathrm{A^{d}}\), \(\mathrm{B^{d}}\), and \(\mathrm{C^{d}}\) correspond to dual sites located at the centers of rhombic tiles, while \(\mathrm{D^{d}}\) and \(\mathrm{E^{d}}\) correspond to dual sites located at the centers of square tiles. Unlike the vertex model, all sites have the same coordination number $z=4$.

As in the vertex model, this five-class description does not make the closure condition exact. The full value of \(\mathcal M_{ab}^{\text{(d)}}(i)\) also depends on how the classes of neighboring dual sites are paired with the bond lengths and dual height differences. This pairing is not always completely fixed by the five classes alone. Thus, we regard the five-class description as a finite-class approximation. The resulting effective Hamiltonian remains finite dimensional and numerically solvable, and its practical construction is described in Appendix~\ref{app:PraCons_M}.

As a result, the problem is reduced to the eigenvalue problem of the five-dimensional effective Hamiltonian $M^{\text{(d)}}$ for each fixed $\kappa'$. For a dual site $i\in S_a$, the $m$-th wave function is given by
\begin{align}
\label{eq:dual_SKK_wavefunction}
\psi_{m,\Lambda}(\mathbf r_i)
=
\Psi_{m,\Lambda}(a)
e^{\kappa' h_{\rm d}(\mathbf r_i)}.
\end{align}
Thus, for an arbitrary real constant \(\kappa'\), the five eigenvectors of \(M^{\text{(d)}}\) generate wave functions of the SKK state for the dual model within the present finite-class description. This construction is the main new application of the present framework. As in the vertex model, this finite-class approximation alone does not guarantee that they are eigenstates of the full tight-binding Hamiltonian.

The construction above is formulated in terms of the dual height function defined from the uncorrected vertex height function \(h(\mathbf r_i)\). In the square approximants, the dual height function is also replaced by the corrected dual height function \(\tilde h_{\rm d}(\mathbf r_i)\), defined in Appendix~\ref{app:periodic_height}. Equivalently, \(\tilde h_{\rm d}(\mathbf r_i)\) is obtained from Eq.~\eqref{eq:height_dual} by replacing \(h(\mathbf r_v)\) with \(\tilde h(\mathbf r_v)\). This correction also affects the finite-class approximation in the dual model for the same reason as in the vertex model. The construction of the finite-dimensional effective Hamiltonian for the dual model on the square approximants, including this effect, is described in Appendix~\ref{app:PraCons_apprM}.

Motivated by the fact that the corresponding construction reproduces the SKK state for the ground state of the vertex model, we focus on the lowest eigenvector \(\psi_{0,\Lambda}(\mathbf r)\) constructed from \(M^{\text{(d)}}\). Furthermore, the imposed boundary condition restricts the values of \(\kappa'\) that can give eigenstates of the approximants. Therefore, in Sec.~\ref{sec:result}, we treat \(\kappa'\) as a variational parameter. Its optimal value is determined by comparing the state constructed from the lowest eigenvector of \(M^{\text{(d)}}\) with the numerical ground state of the full tight-binding Hamiltonian. We then verify whether the resulting state accurately describes the ground state of the dual model.

\section{Results and discussion}
\label{sec:result}
In this section, we numerically examine whether the SKK states constructed in Sec.~\ref{sec:model} describe the ground states of the full tight-binding Hamiltonians in the vertex and dual models. For each square approximant with $n=4,5,6,7$, we first numerically diagonalize the full tight-binding Hamiltonian under periodic boundary conditions. The $m$-th eigenstate obtained from this diagonalization is denoted by $|\psi^{\rm Num}_{m}\rangle$, where $m=0$ corresponds to the ground state and $m=1$ to the first excited state. We compare the numerical ground state $|\psi^{\rm Num}_{0}\rangle$ with the lowest SKK state $\ket{\psi_{\rm{gs}}^{\rm{SKK}}}=\ket{\psi_{0,\Lambda}}$ obtained by solving the effective eigenvalue problems.

The exponential parameters $\kappa$ and $\kappa'$ are determined by maximizing the projection of $|\psi^{\rm SKK}_{\rm gs}\rangle$ onto the two-dimensional low-energy subspace spanned by the numerical ground state and first excited state:
\begin{equation}
\label{eq:F01}
    \mathcal F_{01}
    =
    |\langle \psi^{\rm Num}_{0}
    |\psi^{\rm SKK}_{\rm gs}\rangle|^2
    +
    |\langle \psi^{\rm Num}_{1}
    |\psi^{\rm SKK}_{\rm gs}\rangle|^2 .
\end{equation}
This choice is useful because the lowest two eigenstates can have a small energy separation. After optimizing $\kappa$ or $\kappa'$, we evaluate the individual fidelities
\begin{equation}
F_0 = |\langle \psi^{\rm Num}_{0}|\psi^{\rm SKK}_{\rm gs}\rangle|^2,
\quad
F_1 = |\langle \psi^{\rm Num}_{1}|\psi^{\rm SKK}_{\rm gs}\rangle|^2,
\end{equation}
as well as the energy gap \(\Delta E = E_1-E_0\). Here, $E_0$ and $E_1$ are the eigenenergies of the numerical ground state and the first excited state, respectively, obtained from the full tight-binding Hamiltonian. The quantities $F_0$, $F_1$, \(\mathcal F_{01}\), $\kappa$ or $\kappa'$, and $\Delta E$ are plotted as functions of the tuning parameter. Details of the numerical diagonalization and parameter optimization are provided in Appendix~\ref{app:numerical_procedure}.

The results are presented as follows. We first examine the vertex model in Sec.~\ref{subsec:vertex_numerics} as a benchmark, and then apply the same analysis to the dual model in Sec.~\ref{subsec:dual_numerics}. 

\begin{figure}[b]
\centering
\includegraphics[width=0.47\textwidth]{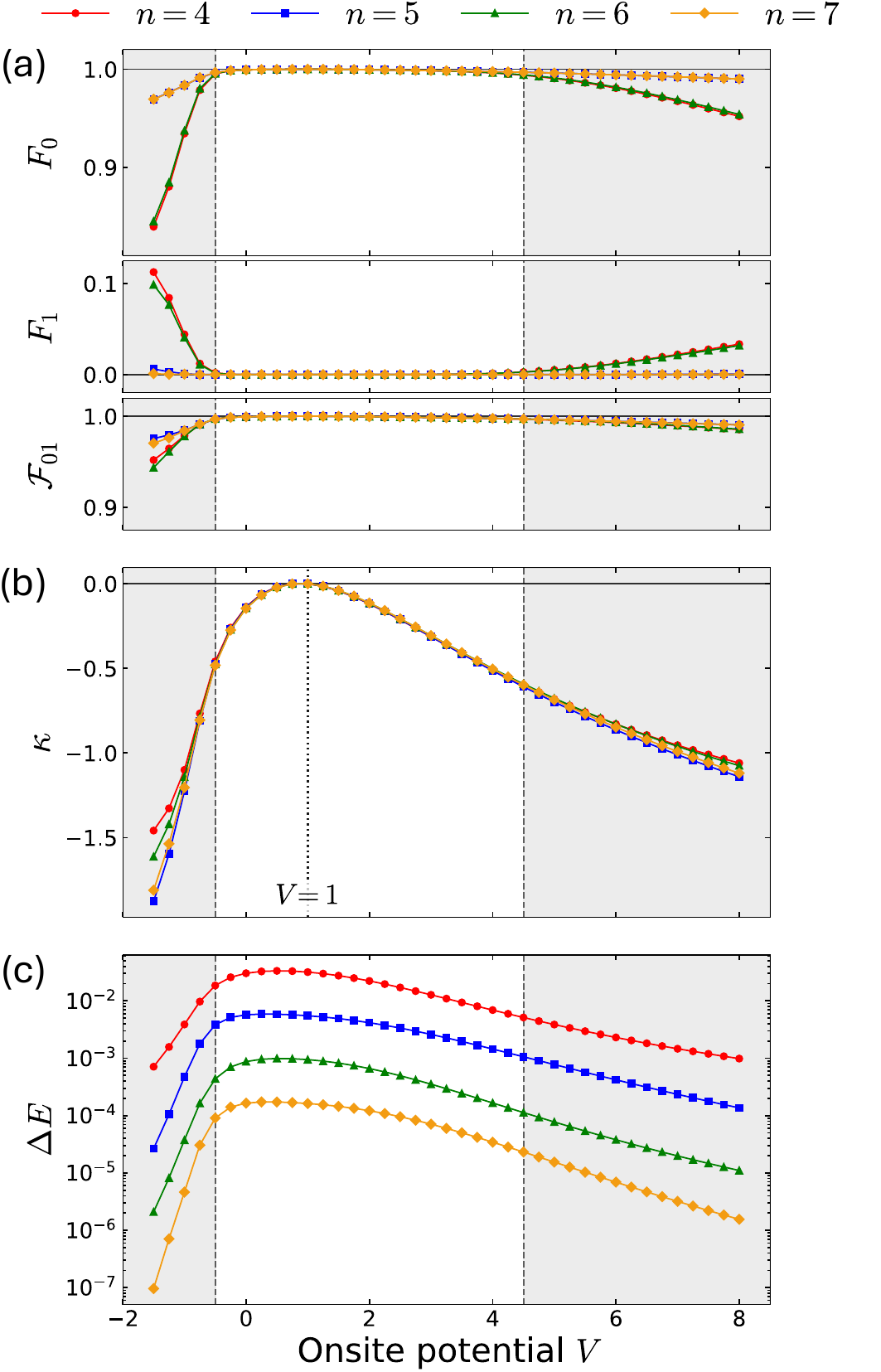}
\caption{Numerical verification of SKK states in the vertex model constructed from square approximants of generations \(n=4,5,6\), and \(7\).
\(|\psi^{\rm Num}_{0}\rangle\) and \(|\psi^{\rm Num}_{1}\rangle\) denote the numerical ground state and first excited state, respectively, and \(|\psi^{\rm SKK}_{\rm gs}\rangle\) denotes the lowest SKK state constructed from the effective Hamiltonian.
(a) Fidelities 
\(F_0=|\langle \psi^{\rm Num}_{0}|\psi^{\rm SKK}_{\rm gs}\rangle|^2\), 
\(F_1=|\langle \psi^{\rm Num}_{1}|\psi^{\rm SKK}_{\rm gs}\rangle|^2\), 
and \(\mathcal F_{01}=F_0+F_1\).
(b) Optimized constant \(\kappa\). The vertical dotted line indicates \(V=1\).
(c) Energy gap \(\Delta E=E_1-E_0\). The shaded regions in all panels indicate parameter regions where the fidelity \(F_0\) is reduced.
}
\label{fig:vertex_graph}
\end{figure}

\subsection{Numerical verification in the vertex model}
\label{subsec:vertex_numerics}
\begin{figure}[b]
\centering
\includegraphics[width=0.5\textwidth]{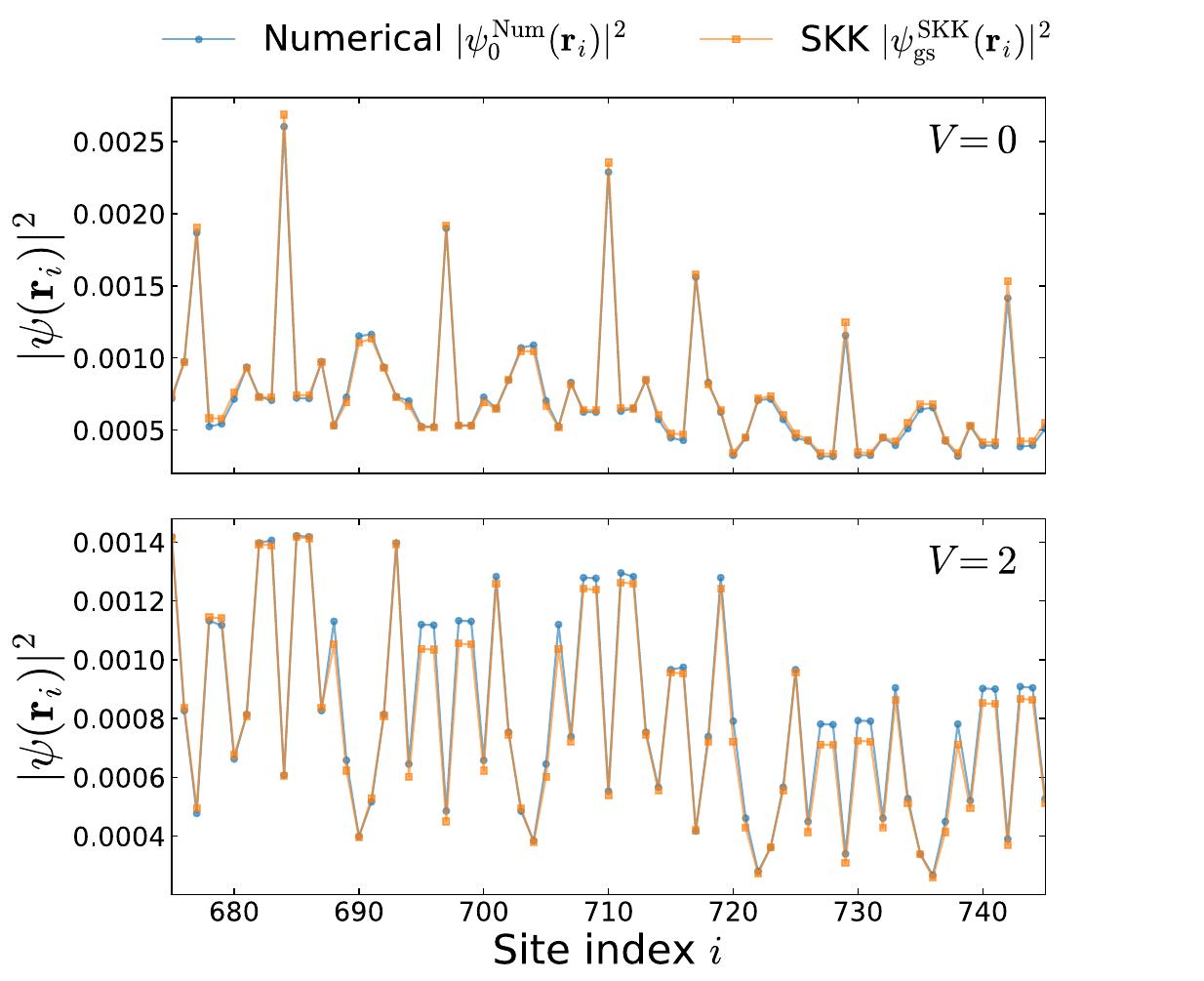}
\caption{Real-space probability distributions of the numerical ground state and the SKK state in the vertex model for the square approximant with \(n=4\). 
The upper and lower panels correspond to \(V=0\) and \(V=2\), respectively. 
The plotted range is \(675\le i\le 745\). 
Blue circles and orange squares denote 
\(|\psi^{\rm Num}_{0}(\mathbf r_i)|^2\) 
and 
\(|\psi^{\rm SKK}_{\rm gs}(\mathbf r_i)|^2\), respectively.
}
\label{fig:wavefunction_vertex}
\end{figure}

Fig.~\ref{fig:vertex_graph} shows the results for the vertex model. The fidelity with the numerical ground state $F_0$ remains close to unity in the region \(-0.5\leq V\leq4.5\), showing that the SKK state constructed within the finite-class approximation accurately describes the ground state in this parameter range. The optimized parameter \(\kappa\) changes continuously with the onsite potential \(V\) and approaches zero near \(V=1\). At $V=1$, the tight-binding Hamiltonian for the vertex model becomes the graph Laplacian, for which the ground state is spatially uniform. The approach of \(\kappa\) to zero means that the factor $e^{\kappa \tilde h(\mathbf r_i)}$ becomes close to unity, consistent with a spatially uniform ground state. This behavior is consistent with the SKK state reported in previous studies~\cite{Kalugin2014,Mace2017}, up to a sign difference arising from the sign convention used for the height function.

The fidelity \(F_0\) starts to decrease in the regions \(V<-0.5\) and \(V>4.5\). This decrease is more clearly seen for approximants with even \(n\), where it is accompanied by an increase in \(F_1\) and by a reduction of the energy gap \(\Delta E\). This behavior can be understood perturbatively: as the energy gap decreases, even a small component of the original Hamiltonian not captured by the finite-class approximation can lead to stronger mixing with the first excited state. In these regions, \(\mathcal F_{01}=F_0+F_1\) remains below unity, suggesting perturbative admixture of higher excited states.

The agreement between the constructed SKK state and the numerical ground state can also be checked directly in real space. Fig.~\ref{fig:wavefunction_vertex} compares the real-space probability distributions
\(|\psi^{\rm Num}_{0}(\mathbf r_i)|^2\) and
\(|\psi^{\rm SKK}_{\rm gs}(\mathbf r_i)|^2\)
for the square approximant with \(n=4\). For both \(V=0\) and \(V=2\), the two distributions agree well throughout the plotted range.

\begin{figure}[b]
\centering
\includegraphics[width=0.47\textwidth]{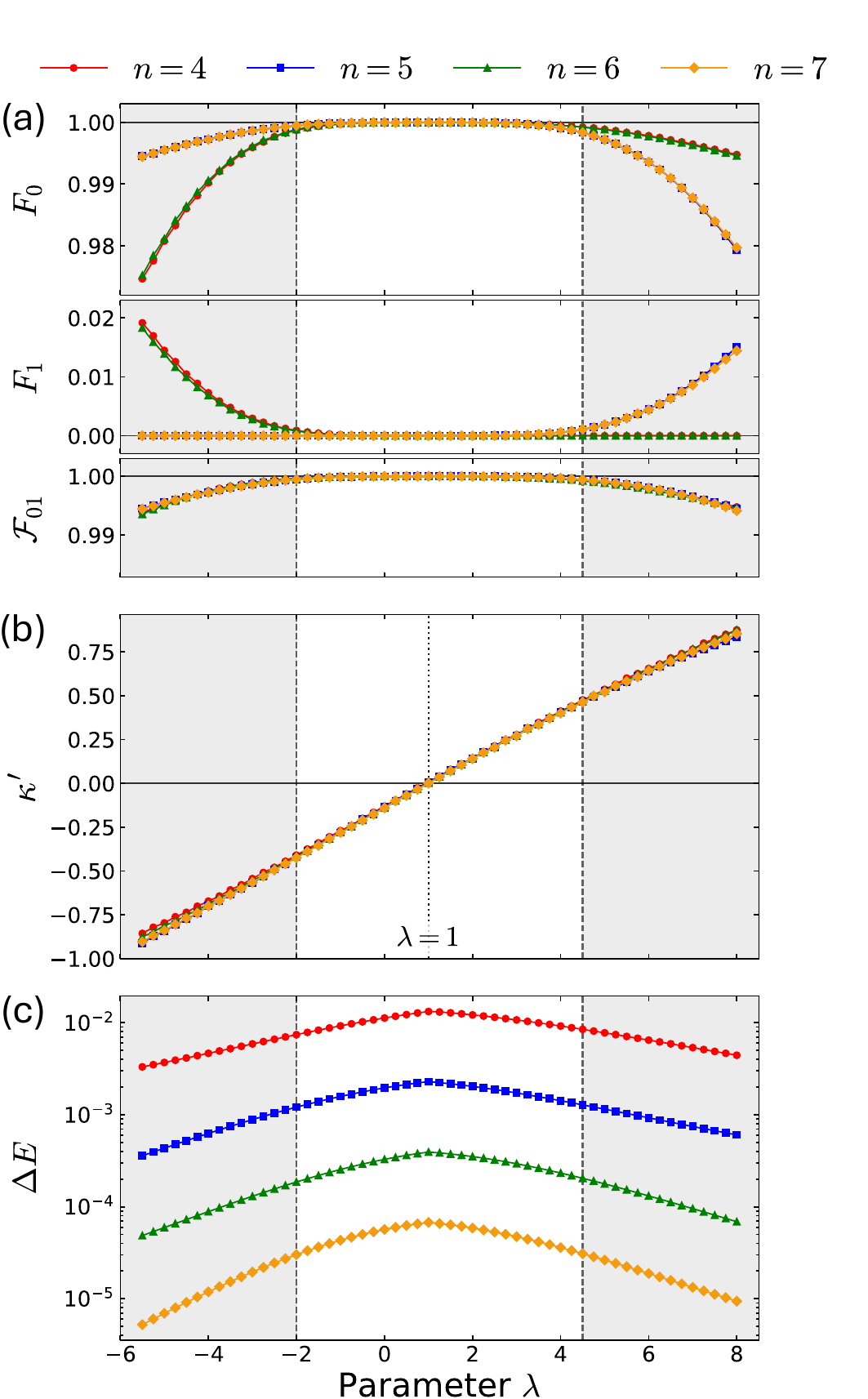}
\caption{
Numerical verification of SKK states in the dual model constructed from square approximants of generations \(n=4,5,6\), and \(7\).
(a) Fidelities 
\(F_0=|\langle \psi^{\rm Num}_{0}|\psi^{\rm SKK}_{\rm gs}\rangle|^2\), 
\(F_1=|\langle \psi^{\rm Num}_{1}|\psi^{\rm SKK}_{\rm gs}\rangle|^2\), 
and \(\mathcal F_{01}=F_0+F_1\).
(b) Optimized constant $\kappa'$. The vertical dotted line indicates $\lambda=1$.
(c) Energy gap $\Delta E=E_1-E_0$. The shaded regions in all panels indicate parameter regions where the fidelity $F_0$ is reduced.
}
\label{fig:dual_graph}
\end{figure}

Taken together, these results show that the SKK state previously reported for the vertex model is obtained constructively within our framework and accurately describes the numerical ground state over a substantial part of the parameter range considered here, where the energy gap does not
become particularly small. The vertex model therefore serves as a benchmark for the present construction.

\subsection{Numerical verification in the dual model}
\label{subsec:dual_numerics}
\begin{figure}[b]
\centering
\includegraphics[width=0.5\textwidth]{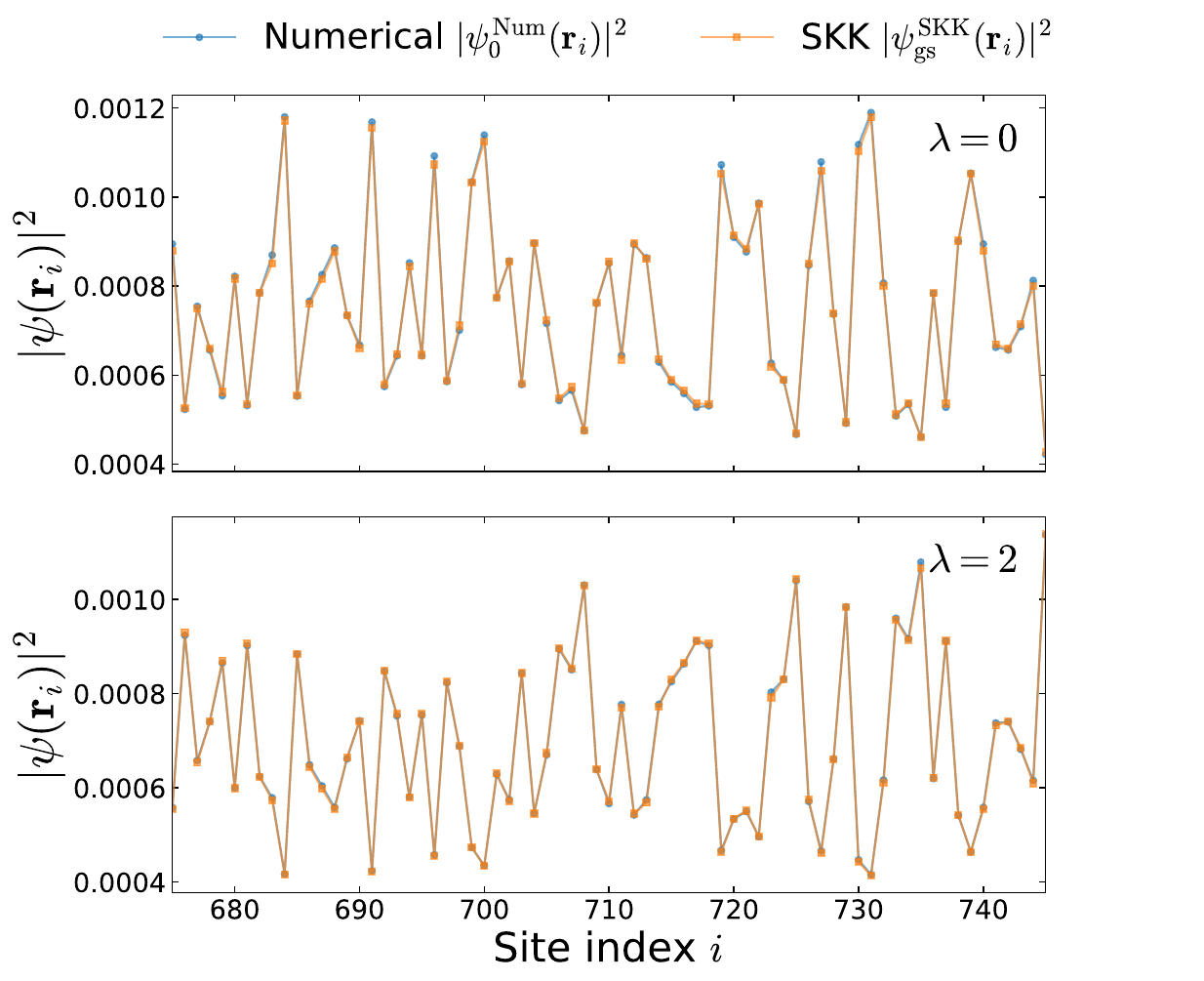}
\caption{Real-space probability distributions of the numerical ground state and the SKK state in the dual model for the square approximant with \(n=4\). The upper and lower panels correspond to \(\lambda=0\) and \(\lambda=2\), respectively. 
The plotted range is \(675\le i\le 745\). 
Blue circles and orange squares denote 
\(|\psi^{\rm Num}_{0}(\mathbf r_i)|^2\) 
and 
\(|\psi^{\rm SKK}_{\rm gs}(\mathbf r_i)|^2\), respectively.
}
\label{fig:wavefunction_dual}
\end{figure}

Fig.~\ref{fig:dual_graph} shows the corresponding results for the dual model. Here, \(\lambda\) controls the ratio \(V_{ij}/t_{ij}=\lambda\) in the Hamiltonian for the dual model. The fidelity with the numerical ground state $F_0$ remains close to unity in the region \(-2\leq \lambda\leq4.5\), showing that the SKK state constructed within the finite-class approximation accurately describes the ground state in this parameter range. 

The optimized parameter $\kappa'$ changes continuously with $\lambda$ and approaches zero near $\lambda=1$. The behavior near $\lambda=1$ can be interpreted in the same way as in the vertex model. At $\lambda=1$, the tight-binding Hamiltonian for the dual model becomes a weighted graph Laplacian, for which the ground state is spatially uniform. As $\kappa'\to 0$, the factor $e^{\kappa' \tilde h_{\rm d}(\mathbf r_i)}$ tends to unity, consistent with a spatially uniform ground state.

The fidelity \(F_0\) starts to decrease in the regions \(\lambda<-2\) and \(\lambda>4.5\). The behavior depends on whether the approximant generation \(n\) is even or odd: the decrease is more clearly seen for even \(n\) in the region \(\lambda<-2\), while it is more clearly seen for odd \(n\) in the region \(\lambda>4.5\). In the cases where the decrease of \(F_0\) is visible, it is accompanied by an increase in \(F_1\) and by a reduction of the energy gap \(\Delta E\). This behavior can be interpreted perturbatively in the same way as in the vertex model. In these regions, \(\mathcal F_{01}=F_0+F_1\) remains below unity, further suggesting perturbative admixture of higher excited states.

The agreement between the constructed SKK state and the numerical ground state can also be checked directly in real space. Fig.~\ref{fig:wavefunction_dual} compares the real-space probability distributions
\(|\psi^{\rm Num}_{0}(\mathbf r_i)|^2\) and
\(|\psi^{\rm SKK}_{\rm gs}(\mathbf r_i)|^2\)
for the square approximant with \(n=4\). For both \(\lambda=0\) and \(\lambda=2\), the two distributions agree well throughout the plotted range.

Taken together, the fidelity and real-space comparisons for the dual model provide evidence that the SKK state obtained within our framework accurately describes the numerical ground state of the dual model over a substantial part of the parameter range considered here, where the energy gap does not become particularly small.

Viewed alongside the results for the vertex model, this finding indicates that SKK states are not restricted to a particular electron-site configuration. This correspondence can be understood by comparing the structures of the two SKK states. In both models, the exponential factor is governed by a height function, and the dual height function is constructed from the vertex height function. The two exponential factors can be regarded as reflecting the underlying quasiperiodicity of the AB tiling, whereas the differences between the two electron-site configurations are encoded in the class-dependent prefactors, which are assigned to site classes determined by local environments. The behavior of these class-dependent prefactors, together with those of the vertex model, is summarized in Appendix~\ref{app:prefactor}.

\section{Conclusion}
\label{sec:conclusion}

In this work, we investigated the construction of SKK states in quasiperiodic tight-binding models from the viewpoint of a generalized Fourier transform and the associated closure condition. We applied this framework to two models on the AB tiling: the vertex model and the dual model. In previous studies~\cite{Kalugin2014,Mace2017}, the SKK state was treated mainly as an ansatz for the vertex model. The present results show that this state is obtained constructively within our framework through the effective eigenvalue problem, and further reveal a corresponding SKK state in the dual model, where an SKK state had not been reported previously. The realization of SKK states in both models suggests that this structure is not tied to a particular choice of electron sites, but can persist across different electron-site configurations sharing the same underlying quasiperiodicity.

The structural relation between the two SKK states is not accidental. The height function used in the dual model is constructed by averaging the vertex height function over the four vertices of each tile. Therefore, the exponential factors $e^{\kappa \tilde h(\mathbf r_i)}$ in the vertex model and $e^{\kappa' \tilde h_d(\mathbf r_i)}$ in the dual model are closely related through the underlying quasiperiodic structure of the tiling. This suggests that the structure of the height function reflects the common quasiperiodicity of the tiling, rather than the particular choice of electron sites. The comparison of the class-dependent prefactors further clarifies the role of local geometry.

In the main parameter regions considered here, the seven-class vertex description and the five-class dual description reproduce the numerical ground states with high fidelity, even though the full closure condition is not imposed. This supports the validity of the finite-class approximation. The deviations from perfect fidelity are also informative. In both models, the decrease in \(F_0\) is accompanied by an increase in \(F_1\) and a reduction of the energy gap. This behavior can be interpreted perturbatively as an enhanced admixture of excited states arising from a residual component of the original Hamiltonian not captured by the finite-class approximation. The fact that \(\mathcal F_{01}<1\) indicates that this admixture is not restricted to the first excited state.

The present construction also suggests a route to searching for generalized Bloch-like states in more general quasiperiodic systems. A natural future direction is to identify suitable exponents \(f_\Lambda(\mathbf r)\) and site classifications \(\{S_a\}\) beyond the AB tiling. In the framework of Kalugin and Katz~\cite{Kalugin2014}, the exponential factor in an SKK state is related to a cohomological object associated with the hull of the quasiperiodic structure. Combining this geometric viewpoint with the closure condition may provide a systematic route to finding SKK states in a broader class of quasiperiodic systems. Such a generalized Bloch-like description may also be useful for studying weak-coupling superconductivity in quasiperiodic systems, which has been explored both experimentally~\cite{Kamiya2018,Tokumoto2024,Terashima2024,Matsudaira2026} and theoretically~\cite{Sakai2017,Araujo2019,Takemori2020,Liu2022, Fukushima2023proceeding,Fukushima2023,Liu2023}. Since the standard momentum-space formulation of Bardeen-Cooper-Schrieffer (BCS) theory in periodic crystals is naturally expressed in terms of Bloch states \cite{Bardeen1957,deGennes1999,Tinkham2004,Kita2015}, constructing Bloch-like states in quasiperiodic systems may provide a useful basis for extending such theoretical descriptions beyond periodic crystals.

\begin{acknowledgments}
N.~T. is supported by JSPS KAKENHI Grant Nos. JP25H01397, JP25H01400, JP24K00585, and JP25K22239.
\end{acknowledgments}
\section*{Data Availability}
The numerical data and codes that support the findings of this study are available from the corresponding author upon reasonable request.
\appendix
\section{Reduction to an effective eigenvalue problem}
\label{app:GFT_SKK}
Here, we demonstrate how the closure condition in Eq.~(\ref{eq:commutator_condition}) reduces the Schrödinger equation of the Hamiltonian $\hat{\mathscr{H}}$ to the effective eigenvalue problem in Eq.~(\ref{eq:FiniteDim_EP}). Assuming the closure condition is satisfied, the Hamiltonian $\hat{\mathscr{H}}$ acts within the invariant subspace spanned by $\{\hat{c}_{a,\Lambda}^\dagger|0\rangle\}_a$. We define an arbitrary state $|\Phi_{\Lambda}\rangle$ within this subspace as
\begin{align}
\label{eq:app_state}
|\Phi_{\Lambda}\rangle = \sum_b \Psi_{\Lambda}(b) \hat{c}_{b,\Lambda}^\dagger |0\rangle.
\end{align}
Operating on this state with the Hamiltonian and using the closure condition, we obtain
\begin{align}
\hat{\mathscr{H}}|\Phi_{\Lambda}\rangle &= \sum_b \Psi_{\Lambda}(b) [\hat{\mathscr{H}},\hat{c}_{b,\Lambda}^\dagger]|0\rangle \nonumber \\
&= \sum_b \Psi_{\Lambda}(b) \sum_a M_{ab} \hat{c}_{a,\Lambda}^\dagger|0\rangle \nonumber \\
&= \sum_a \left( \sum_b M_{ab} \Psi_{\Lambda}(b) \right) \hat{c}_{a,\Lambda}^\dagger|0\rangle.
\end{align}
Imposing the Schrödinger equation $\hat{\mathscr{H}}|\Phi_{\Lambda}\rangle = E|\Phi_{\Lambda}\rangle$ yields
\begin{align}
\sum_a \left( \sum_b M_{ab} \Psi_{\Lambda}(b) \right) \hat{c}_{a,\Lambda}^\dagger|0\rangle = E \sum_a \Psi_{\Lambda}(a) \hat{c}_{a,\Lambda}^\dagger |0\rangle.
\end{align}
Since the states $\{\hat{c}_{a,\Lambda}^\dagger|0\rangle\}_a$ are linearly independent, the coefficients for each label $a$ must be equal on both sides, leading to
\begin{align}
\sum_b M_{ab} \Psi_{\Lambda}(b) = E \Psi_{\Lambda}(a).
\end{align}
This confirms that the Schrödinger equation of the Hamiltonian $\hat{\mathscr{H}}$ is equivalent to the eigenvalue problem governed by the effective Hamiltonian.

\section{Closure conditions for the Ammann-Beenker tiling}
\label{app:Cc_AB}
Here, we show that the closure condition for the vertex and dual models can be expressed as the requirement that the quantities $\mathcal M_{ab}^{\text{(v)}}(i)$ and $\mathcal M_{ab}^{\text{(d)}}(i)$ introduced in Eqs.~(\ref{eq:SKK_condition_vertex_local}) and (\ref{eq:SKK_condition_dual_local}) be independent of the particular choice of the site $i$ within each class $S_a$.
We begin by considering a general tight-binding Hamiltonian expressed as
\begin{align}
\hat{\mathscr H}
=
\sum_{i,j}H_{ij}\hat c_i^\dagger\hat c_j.
\end{align}
To evaluate the general closure condition in Eq.~(\ref{eq:commutator_condition}), we first calculate the commutator of the Hamiltonian, expanding the commutator to obtain
\begin{align}
[\hat{\mathscr{H}}, \hat{c}_i^\dagger] &= \sum_{j,k} H_{jk} [\hat{c}_j^\dagger \hat{c}_k, \hat{c}_i^\dagger] \nonumber \\
&= \sum_j H_{ji} \hat{c}_j^\dagger.
\end{align}

We then apply this expression to the transformed creation operator $\hat{c}_{a,\Lambda}^\dagger$ defined in Eq.~(\ref{eq:GeneralizedFourier2}). Multiplying by the generalized Fourier factor $e^{f_\Lambda(\mathbf{r}_i)}$ and summing over the subset $S_a$ expands the commutator as
\begin{align}
[\hat{\mathscr{H}}, \hat{c}_{a,\Lambda}^\dagger] 
&= \sum_j \left(\sum_{i \in S_a} H_{ji} e^{f_\Lambda(\mathbf{r}_i)} \right)\hat{c}_j^\dagger.
\label{eq:app_comm_general}
\end{align}

We now compare the explicit commutator in Eq.~(\ref{eq:app_comm_general}) with the right-hand side of the general closure condition in Eq.~(\ref{eq:commutator_condition}). The latter can be expanded as
\begin{align}
\label{eq:app_comm_target}
\sum_b M_{ba} \hat{c}_{b,\Lambda}^\dagger &= \sum_b \sum_{j \in S_b} M_{ba} e^{f_\Lambda(\mathbf{r}_j)} \hat{c}_j^\dagger.
\end{align}
Because the creation operators $\hat{c}_j^\dagger$ are linearly independent, their coefficients in Eqs.~\eqref{eq:app_comm_general} and \eqref{eq:app_comm_target} must be identical. Let \(\tilde a\) denote the class containing site \(j\), so that $j\in S_{\tilde a}$. Then the coefficient of $\hat{c}_j^\dagger$ in Eq.~(\ref{eq:app_comm_target}) is $M_{\tilde a a}e^{f_\Lambda(\mathbf{r}_j)}$. Therefore, comparing these coefficients yields
\begin{align}
\sum_{i \in S_a} H_{ji} e^{f_\Lambda(\mathbf{r}_i)}
=
M_{\tilde a a} e^{f_\Lambda(\mathbf{r}_j)}
\qquad
(j\in S_{\tilde a}).
\end{align}
This condition can be rewritten as
\begin{align}
M_{ab}
=
\sum_{j\in S_b}
H_{ij}
e^{f_\Lambda(\mathbf r_j)-f_\Lambda(\mathbf r_i)}
\qquad
(i\in S_a).
\label{eq:app_general_M}
\end{align}
Since the matrix elements $M_{ab}$ in the closure condition are labeled only by the site classes, the closure condition requires the right-hand side of Eq.~\eqref{eq:app_general_M} to be independent of the particular choice of the site $i$ within the class $S_a$. To express this requirement explicitly, for each pair of classes $(a,b)$, we regard the right-hand side as a function on the set $S_a$:
\begin{align}
\mathcal M_{ab}(i)
\equiv
\sum_{j\in S_b}
H_{ij}
e^{f_\Lambda(\mathbf r_j)-f_\Lambda(\mathbf r_i)}
\qquad
(i\in S_a).
\label{eq:app_local_M}
\end{align}
The closure condition is the requirement that this quantity be independent of the particular choice of $i$ within the same class $S_a$. 
When this condition is satisfied, we can write
\begin{align}
\label{eq:M_c}
\mathcal M_{ab}(i)=M_{ab}
\qquad
\text{for all } i\in S_a .
\end{align}

Applying Eq.~\eqref{eq:app_local_M} to the Hamiltonian matrix for the vertex model
\begin{align}
H_{ij}^{\mathrm{(v)}}=-t\delta_{\langle i,j\rangle}+Vz_i\delta_{ij}
\end{align}
gives Eq.~\eqref{eq:SKK_condition_vertex_local}. Similarly, applying it to the Hamiltonian matrix for the dual model
\begin{align}
H_{ij}^{\mathrm{(d)}}=
-t_{ij}\delta_{\langle i,j\rangle}
+
\left(\sum_{k\in N(i)}V_{ik}\right)\delta_{ij}
\end{align}
gives Eq.~\eqref{eq:SKK_condition_dual_local}. Together with
Eq.~\eqref{eq:M_c}, these expressions lead to the closure
conditions in Eqs.~\eqref{eq:SKK_condition_vertex} and
\eqref{eq:SKK_condition_dual}.

\section{Finite-class approximation to the closure condition}
\label{app:PraCons_M}
Here, we describe how the finite-dimensional effective Hamiltonians used in Secs.~\ref{subsubsec:vertex_skk} and \ref{subsubsec:dual_skk} are constructed. Within the finite-class approximations, the local closure quantities 
\(\mathcal M_{ab}^{\text{(v)}}(i)\) and 
\(\mathcal M_{ab}^{\text{(d)}}(i)\) can still depend weakly on the choice of the site \(i\in S_a\). We therefore define the effective Hamiltonian by selecting a representative site \(i_a^{\rm rep}\) for each class \(S_a\), as described below.

\subsection{Vertex model}
In the vertex model, the remaining dependence of \(\mathcal M_{ab}^{\text{(v)}}(i)\) on the particular site \(i\in S_a\) comes from how the classes of neighboring sites are paired with the height differences on the corresponding bonds. For each site \(i\in S_a\), we therefore define
\begin{align}
\label{eq:Nsabi_v}
\mathcal{N}_{ab}^{\sigma}(i)
=
\#\left\{
j\in N(i)\cap S_b
\ \middle|\ 
\Delta h_{ij}=\sigma
\right\}.
\end{align}
Here, \(\#\{\cdots\}\) denotes the number of elements in the set, and $\Delta h_{ij} = h(\mathbf r_j)-h(\mathbf r_i)$.
For each site \(i\in S_a\), the set of integers
\begin{align}
\mathcal P_a^{\mathrm{(v)}}(i) = \left\{ \mathcal{N}_{ab}^{\sigma}(i) \right\}_{b,\,\sigma=\pm1}
\end{align}
specifies the pairing of the local data entering the closure condition. For each class \(S_a\), we choose, as the representative site \(i_a^{\mathrm{rep}}\), one of the sites whose pattern \(\mathcal P_a^{\mathrm{(v)}}(i)\) occurs most frequently among the sites in \(S_a\). If several of the most frequent patterns are tied, we choose one site from any of them. The matrix elements of the effective Hamiltonian are then defined as
\begin{align} M_{ab}^{\mathrm{(v)}}  = \mathcal M_{ab}^{\mathrm{(v)}}(i_a^{\mathrm{rep}}). 
\end{align}

\subsection{Dual model}
In the dual model, the remaining dependence of \(\mathcal M_{ab}^{\text{(d)}}(i)\) on the particular site \(i\in S_a\) comes from how the classes of neighboring sites are paired with the bond lengths and the dual height differences on the corresponding bonds. For each site \(i\in S_a\), we therefore define
\begin{align}
\label{eq:Nsabi_d}
\mathcal{N}_{ab}^{\rho\eta}(i)
=
\#\left\{
j\in N(i)\cap S_b
\ \middle|\ 
(\rho_{ij},\Delta h^{\mathrm d}_{ij})=(\rho,\eta)
\right\}.
\end{align}
Here, \(\rho_{ij}=|\mathbf r_j-\mathbf r_i|\) denotes the bond length, and $\Delta h^{\mathrm d}_{ij} = h_{\mathrm d}(\mathbf r_j)-h_{\mathrm d}(\mathbf r_i)$. The bond length \(\rho_{ij}\) is used to distinguish the three bond types: square-square, square-rhombus, and rhombus-rhombus. The dual height difference across a nearest-neighbor bond takes only the values
\begin{align}
\Delta h^{\mathrm d}_{ij} \in \left\{ -\frac{1}{2},0,\frac{1}{2} \right\}.
\end{align}
Thus, \(\rho\) runs over the three bond types, while \(\eta\) runs over \(\{-1/2,0,1/2\}\). For each site \(i\in S_a\), the set of integers
\begin{align}
\mathcal P_a^{\mathrm{(d)}}(i) = \left\{ \mathcal{N}_{ab}^{\rho\eta}(i) \right\}_{b,\,\rho,\,\eta}
\end{align}
specifies the local data entering the closure condition. For each class \(S_a\), we choose, as the representative site \(i_a^{\mathrm{rep}}\), one of the sites whose pattern \(\mathcal P_a^{\mathrm{(d)}}(i)\) occurs most frequently among the sites in \(S_a\). If several of the most frequent patterns are tied, we choose one site from any of them. The matrix elements of the effective Hamiltonian are then defined as
\begin{align}
M_{ab}^{\mathrm{(d)}} = \mathcal M_{ab}^{\mathrm{(d)}}(i_a^{\mathrm{rep}}).
\end{align}

\section{Construction for the square approximants}
\subsection{Periodic height function for the square approximants}
\label{app:periodic_height}
As discussed in Secs.~\ref{subsubsec:vertex_skk} and \ref{subsubsec:dual_skk}, the height function used for the square approximants must be made compatible with the imposed periodic boundary conditions. Here, we describe the explicit correction used in the numerical calculations. As discussed in Sec.~\ref{subsec:ABtiling}, the square approximants are generated by replacing the exact transformation matrix $C$ with its rational approximation $C_n$ in the cut-and-project method. Geometrically, this approximation tilts the projection planes $\mathcal{S}_{||}$ and $\mathcal{S}_{\perp}$ in the four-dimensional space. Since this change is a simple tilt, we expect that the resulting shift in the height function can be corrected by a linear function of the spatial coordinates $\mathbf{r}_i = (x_i, y_i)$. Following this expectation, we introduce a linear correction term
\begin{equation}
\label{eq:hsurf}
h_{\text{surf}}(\mathbf{r}_i) = \frac{x_i - y_i}{L} + c
\end{equation}
where \(L\) is the side length of the unit cell and \(c\) is an arbitrary constant. The corrected height function is then defined as
\begin{equation}
\label{eq:hchilde}
\tilde{h}(\mathbf{r}_i) = h(\mathbf{r}_i) + h_{\text{surf}}(\mathbf{r}_i).
\end{equation}
The linear part of \(h_{\mathrm{surf}}\) removes the mismatch of the height function across the boundaries of the unit cell. The constant \(c\) fixes only the overall additive constant of the height function and does not affect the normalized SKK state. In the present convention, we set \(c=1\). Thus, \(\tilde h(\mathbf r_i)\) is compatible with the periodic boundary conditions, and we obtain a well-defined generalized Fourier factor \(e^{\kappa \tilde h(\mathbf r_i)}\). In the dual model, we define the height function on dual sites by applying Eq.~(\ref{eq:height_dual}) to the periodic vertex height:
\begin{align}
\tilde h_{\rm d}(\mathbf r_i)
=
\frac{1}{4}
\sum_{v\in\mathcal T(i)}
\tilde h(\mathbf r_v).
\end{align}
The corresponding generalized Fourier factor is \(e^{\kappa' \tilde h_{\rm d}(\mathbf r_i)}\).

\subsection{Finite-class approximation with periodic height functions}
\label{app:PraCons_apprM}

As described in Secs.~\ref{subsubsec:vertex_skk} and \ref{subsubsec:dual_skk}, we use the corrected height functions \(\tilde h(\mathbf r_i)\) and \(\tilde h_{\rm d}(\mathbf r_i)\) defined above. Because these corrected height functions contain a small linear correction, their nearest-neighbor differences $\tilde h(\mathbf r_j)-\tilde h(\mathbf r_i)$ and $\tilde h_{\rm d}(\mathbf r_j)-\tilde h_{\rm d}(\mathbf r_i)$
do not take exactly the discrete values of the quasiperiodic AB tiling. Therefore, when identifying the representative sites $i^{\mathrm{rep}}_a$, we assign the corrected height differences to the corresponding discrete values.

For the vertex model, we define
\begin{align}
\Delta \tilde h_{ij} = \mathcal R_{\mathrm v} \left( \tilde h(\mathbf r_j)-\tilde h(\mathbf r_i) \right),
\end{align}
where
\begin{align}
\mathcal R_{\mathrm v}(x) = 
\begin{cases} 
+1, & x>0,\\ 
-1, & x<0. 
\end{cases}
\end{align}
In the present construction, no zero value appears for nearest-neighbor bonds. Thus, \(\Delta \tilde h_{ij}\) only takes the values \(+1\) and \(-1\), and is used as \(\Delta h_{ij}\) in Eq.~\eqref{eq:Nsabi_v}.

For the dual model, we similarly define
\begin{align}
\Delta \tilde h^{\mathrm d}_{ij} = \mathcal R_{\mathrm d} \left( \tilde h_{\mathrm d}(\mathbf r_j)-\tilde h_{\mathrm d}(\mathbf r_i) \right),
\end{align}
where \(\mathcal R_{\mathrm d}\) assigns its argument to the nearest value in
$\left\{ -1/2, 0, 1/2 \right\}.$
Thus,
\begin{align}
\Delta \tilde h^{\mathrm d}_{ij} \in \left\{ -\frac{1}{2},0,\frac{1}{2} \right\}.
\end{align}
This rounded value is used as \(\Delta h^{\mathrm d}_{ij}\) in Eq.~\eqref{eq:Nsabi_d}.


\section{Numerical procedure} \label{app:numerical_procedure}

Here, we provide details of the numerical calculations used to obtain the results in Sec.~\ref{sec:result}. The low-energy eigenstates of the full tight-binding Hamiltonians were obtained using an implicitly restarted Lanczos method. The parameters \(\kappa\) and \(\kappa'\) were optimized over the interval \([-8,8]\) by evaluating Eq.~\eqref{eq:F01} on a uniform grid of 161 points and then performing a bounded scalar optimization around the best grid point. In the dual model, when several nearly equivalent maxima occurred, we selected the one closest in \(\kappa'\) to the optimum at the preceding value of \(\lambda\).

We next describe the numerical treatment of the effective Hamiltonians. Since the effective Hamiltonians obtained under the finite-class approximation are not guaranteed to be Hermitian, we selected the right eigenvector corresponding to the eigenvalue with the smallest real part. We also monitored the imaginary parts of the effective eigenvalues as a diagnostic of the approximation. For square approximants of generations \(n=4,5,6,7\), the imaginary part of the selected eigenvalue remained negligible within numerical precision throughout the parameter ranges examined in this work.

\section{Comparison of the prefactors}
\label{app:prefactor}
\begin{figure}[t]
\centering
\includegraphics[width=0.48\textwidth]{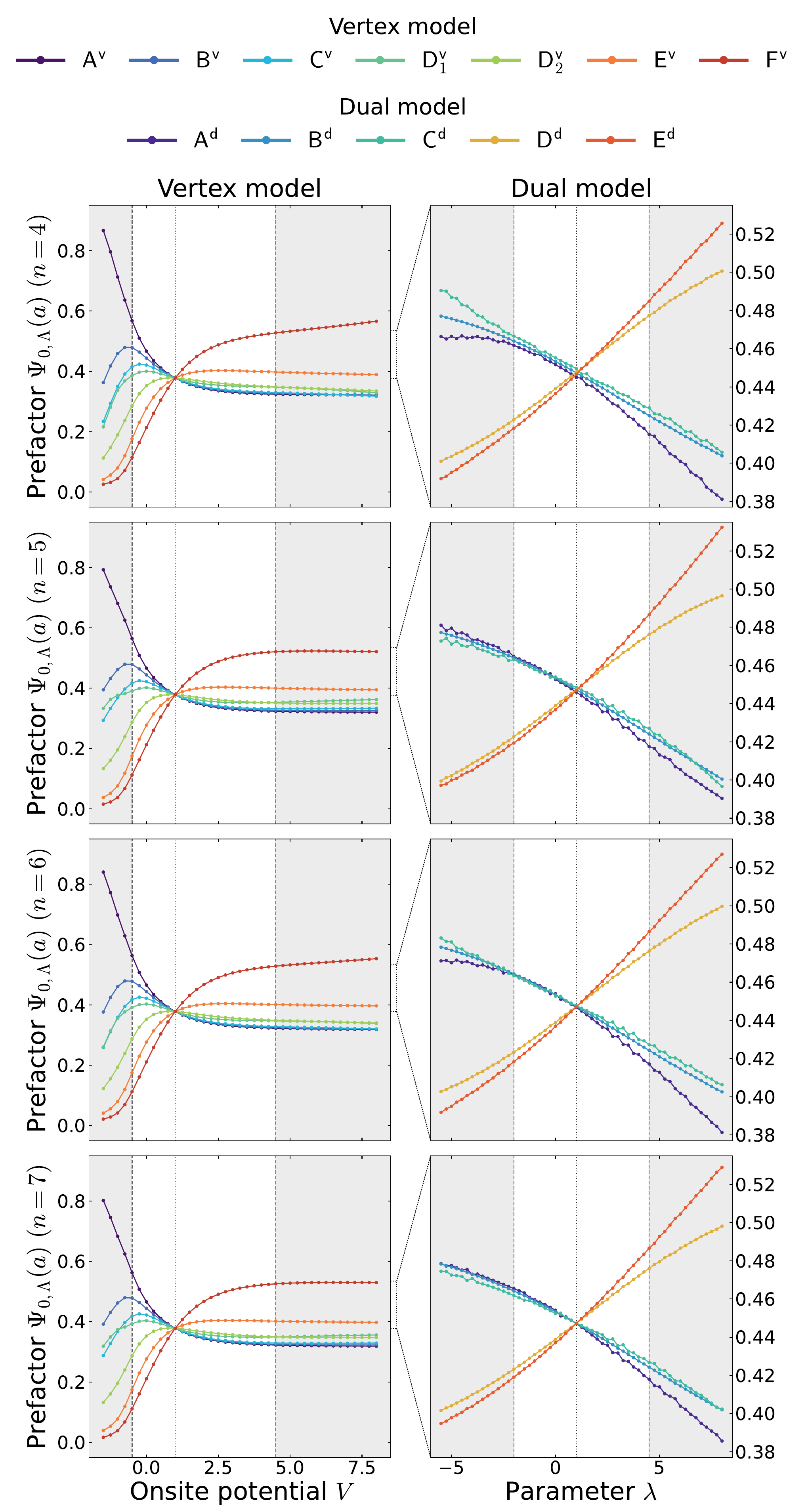}
\caption{
Comparison of the class-dependent prefactors \(\Psi_{0,\Lambda}(a)\) in the vertex and dual models. 
The left and right panels show the vertex and dual results, respectively, with the dual results plotted on a narrower vertical scale. 
The prefactors are normalized as \(\sum_a |\Psi_{0,\Lambda}(a)|^2=1\). 
The vertical dotted lines indicate \(V=1\) and \(\lambda=1\), and the shaded regions show where \(F_0\) is reduced.
}
\label{fig:prefactor_graph}
\end{figure}

Here, we show the class-dependent prefactors \(\Psi_{0,\Lambda}(a)\) obtained from the effective eigenvalue problems for the vertex and dual models. These prefactors are the components of the lowest eigenvector of the corresponding effective Hamiltonian and determine the class-dependent part of the SKK state in the finite-class description. Fig.~\ref{fig:prefactor_graph} shows the results for the square approximants with \(n=4,5,6,7\).

In the vertex model, the prefactors show almost no visible dependence on the approximant generation. Around \(V=1\), the components become nearly equal, consistent with the spatially uniform ground state of the graph Laplacian. Away from this point, the components show a clear class dependence. In particular, the \(V=0\) data show a class-dependent prefactor structure consistent with the previously reported SKK state~\cite{Kalugin2014,Mace2017}.

In the dual model, the overall parameter dependence is similar for \(n=4,5,6,7\), with only a small odd-even difference for \(\lambda<1\). The five prefactor components vary within a much narrower range than those in the vertex model. This weaker class dependence is consistent with the fact that the coordination number is uniform in the dual model. Around \(\lambda=1\), the components become nearly equal, as expected from the spatially uniform ground state of the weighted graph Laplacian.

\input{main.bbl}
\end{document}

%% file: main.bbl
%